\documentclass[11pt,a4paper]{article}

\usepackage{jheppub}
\usepackage{amsfonts}
\usepackage{epsfig}
\usepackage{amssymb}
\usepackage{amsmath}
\usepackage{graphicx}
\usepackage{yfonts}

\newcommand{\be}{\begin{equation}}
\newcommand{\ee}{\end{equation}}
\newcommand{\bea}{\begin{eqnarray}}
\newcommand{\eea}{\end{eqnarray}}

\newcommand{\sac}{\, , \qquad}
\newcommand{\mt}[1]{\textrm{\tiny #1}}
\newcommand{\jem}{J^\mt{EM}}

\newcommand{\vev}[1]{\langle #1\rangle}
\newcommand{\wn}{{\textswab{w}}}
\def\nc {N_\mt{c}}
\def\nf {N_\mt{f}}

\def\gym {g_\mt{YM}}

\title{Brighter Branes,  \\ enhancement of photon production by strong magnetic fields in the gauge/gravity correspondence}
\author[a]{Gustavo Arciniega}
\author[a]{Francisco Nettel}
\author[a]{Patricia Ortega}
\author[a]{ and Leonardo Pati\~no}
\affiliation[a]{Departamento de F\'\i sica, Facultad de Ciencias, Universidad Nacional Aut\'onoma de M\'exico, A.P. 50-542, M\'exico D.F. 04510, M\'exico}

\date{\today}

\abstract{
We use the gauge/gravity correspondence to calculate the rate of photon production in a strongly coupled ${\cal N}=4$ plasma in the presence of an intense magnetic field. We start by constructing a family of back reacted geometries that include the black D3-brane solution, as a smooth limiting case for $B=0$, and extends to backgrounds with an arbitrarily large constant magnetic field. This family provides the gravitational dual of a field theory in the presence of a very strong magnetic field which intensity can be fixed as desired and allows us to study its effect on the photon production of a quark-gluon plasma. The inclusion of perturbations in the electromagnetic field on these backgrounds is consistent only if the metric is perturbed as well, so we use methods developed to treat operator mixing to manage these general perturbations. Our results show a clear enhancement of photon production with a significant anisotropy, which, in qualitative agreement with the experiments of heavy ion collisions, is particularly noticeable for low P.}


\emailAdd{gustavo.arciniega@gmail.com} 
\emailAdd{fnettel@ciencias.unam.mx} 
\emailAdd{patmxm@gmail.com} 
\emailAdd{leopj@ciencias.unam.mx}

\begin{document}  

\begin{flushright}
MAD-TH-12-07
\end{flushright}

\maketitle
\setlength{\parskip}{8pt}


\section{Introduction}

The measurement of direct photons is an important test for any model trying to describe the physics of the state created in pp or heavy-ion collisions like those at RHIC \cite{rhic,rhic2} and LHC \cite{lhc}. Due to the weakness of the electromagnetic coupling and the small extension in space of the quark-gluon plasma (QGP) produced in such collisions, the photons propagate in a medium where their re-scattering is unlikely, therefore representing an excellent probe of the conditions at the emission point \cite{photon,photon1}.

Perturbative calculations of the production rate of photons produced in a plasma are possible at small $\alpha_\mt{s}$, however there is a sound indication that the QGP produced at RHIC and LHC remains strongly coupled \cite{fluid,fluid2}, and the predictions obtained by this method are expected to have a limited applicability. This non-perturbative nature of the plasma provides a motivation for the application of the gauge/gravity correspondence \cite{duality,duality2,duality3}\footnote{See \cite{review} for a review of  applications to the QGP.} to this system. In spite of the success of the AdS/CFT holography program, the exact gravitational setup for QCD has not been found yet. Nevertheless, the study of generic theories, such as finite temperature ${\cal N}=4$ super Yang-Mills (SYM), may lead to valid qualitative predictions about the properties of the dynamics of the plasma at strong coupling. In fact, some features are universal and independent of the model, e.g. the shear viscosity to entropy density ratio of the plasma \cite{Kovtun:2004de}.\footnote{At zero temperature QCD is a confining theory while SYM ${\cal N}=4$ is a conformal and supersymmetric theory. At temperatures such as those produced at RHIC and LHC these differences are less apparent. \cite{soup}}    
 
To test the appropriateness of some backgrounds to study the QGP via the duality, we can look for measurements that are not accounted for by the perturbative methods and study them under the gauge/gravity correspondence.

Such is the case of the direct photon enhancement reported by the measurement of the direct photon transverse momentum ($\mathrm{p}_T$) spectrum in Pb-Pb collisions at $\sqrt{S_{NN}}$ = 2.76 TeV with data taken by the ALICE experiment \cite{Wilde:2012wc}, where a clear direct-photon signal for 0-40\% most central collisions below 4 GeV/c was reported. Above this value of $\mathrm{p}_T$ the results are in agreement with pQCD predictions. For a baseline measurement, the analysis was performed for proton-proton collisions at $\sqrt{s}$= 7 TeV, and for peripheral (40-80\%) Pb-Pb collisions. Both results showed no low $\mathrm{p}_T$ direct-photon signal and were in agreement with pQCD calculations.

Another interesting report \cite{Adare:2011zr} is that of the anisotropy determined when the second Fourier component v2 of the azimuthal anisotropy with respect to the reaction plane was measured for direct photons at mid rapidity and transverse momentum of 1–13 GeV/c in Au+Au collisions at $\sqrt{S_{NN}}$= 200 GeV with the PHENIX detector at the Relativistic Heavy Ion Collider. For $\mathrm{p}_T< 4$ GeV/c they found a direct photon v2, comparable to that of hadrons, that was considerably under predicted by the model calculations for thermal photons in this kinematic region.

It was suggested \cite{Basar:2012bp} that a mechanism emerging from the conformal anomaly of QCD$\times$QED and the existence of strong (electro)magnetic fields in heavy ion collisions could account for the unexpected results.

The objective of the present work is to study photon production in a field theory in the presence of an intense magnetic field using the gauge/gravity correspondence. By considering arbitrary intensities of the magnetic field we will show that our results confirm the enhancement of photon production, and by performing the calculations for propagation directions along and perpendicular to the magnetic field, the existence of anisotropy with respect to the reaction plane will be reported. 

The first study of photon emission in a ${\cal N}=4$ plasma at strong coupling by means of the AdS/CFT holography was presented in \cite{CaronHuot:2006te}, using massless quarks in the adjoint representation coupled to the photons. Later \cite{Parnachev:2006ev} and \cite{Mateos:2007yp} used massless and massive quarks in the fundamental representation as the electrically charged particles. A `soft wall' AdS/QCD model was considered to study photon production in \cite{Atmaja:2008mt,Bu:2012zza} and in \cite{Jo:2010sg} a electrically charged black hole background was used for the same purpose. In \cite{corr1,corr2,corr3} string corrections to supergravity were implemented to calculate the electric conductivity of the plasma and the spectral density of the produced photons.

All these works use, for the gravitational setup, geometries that are isotropic in the directions corresponding to the gauge theory. It seems that a relevant property of the QGP produced at the collisioners is its anisotropy. Because of the different pressures along the beam and the transverse directions at the very beginning of the evolution of the plasma, it is expected that the system develops this measurable initial anisotropy \cite{anis1,anis2}. Using the geometric dual setup given by \cite{Janik:2008tc}, the first analysis of the electromagnetic signatures in a strongly coupled anisotropic plasma was done in \cite{Rebhan:2011ke}. This model presents a naked singularity, and although it is possible to impose infalling boundary conditions and to define retarded correlators, this is considered a not convenient feature.

The computation of photon production in a strongly coupled anisotropic plasma of massless quarks using as gravitational dual the type IIB supergravity solution that was discovered in \cite{Mateos:2011ix,Mateos:2011tv}\footnote{In the work mentioned here, the geometry found in \cite{ALT} is generalized for finite temperature, and it has received a good amount of attention as can be seen in the studies that consider this anisotropy to investigate its impact in the shear viscosity to entropy density ratio \cite{rebhan_viscosity,mamo}, the drag force on a heavy quark \cite{Chernicoff:2012iq,giataganas}, the energy lost by a rotating quark \cite{fadafan}, the stopping distance of a light probe \cite{stopping}, the jet quenching parameter of the medium \cite{giataganas,Rebhan:2012bw,jet}, the potential between a quark and antiquark pair, both static \cite{giataganas,Rebhan:2012bw,Chernicoff:2012bu,indians} and in a plasma wind \cite{Chernicoff:2012bu}, including its imaginary part \cite{Fadafan:2013bva}, Langevin diffusion and Brownian motion \cite{langevin,langevin2} and chiral symmetry breaking \cite{Ali-Akbari:2013txa}.} was studied in \cite{Patino:2012py} for arbitrary values of the anisotropy and the direction of propagation of the photons. By embedding flavour branes in the same background \cite{Mateos:2011ix,Mateos:2011tv}, the case for the anisotropic plasma of massive quarks was studied in \cite{Wu:2013qja}, where the photons were considered to propagate either parallel or perpendicular to the ``anisotropic direction", as it was called in \cite{Mateos:2011ix,Mateos:2011tv}. In this last work \cite{Wu:2013qja} a magnetic field was introduced, and its back reaction on the embedding of the branes was considered, but not the back reaction of the background nor the perturbation on the embedding of the D7 induced by the perturbations of the electromagnetic field necessary to study photon production, that as will be apparent below, these perturbations on the embedding are mandatory to keep the calculation consistent. The effect on photon production and electric conductivity of the background magnetic field introduced in this way was successfully reported in \cite{Wu:2013qja}.

As pointed out in \cite{D'Hoker:2009mm}, the back reaction, on the background metric, of an intense magnetic field as the one that has been suggested to source the photon enhancement and anisotropy v2, has to be considered and it leads to a different geometry for the proper dual of a field theory in the presence of such a background field.

In \cite{Mamo2} photon and dilepton production was studied in the presence of a magnetic field using the RG flow equations in the same background that we will use in the present work. The flow equations used in that work do not include the metric perturbations that, as mentioned earlier and will be clarified below, are mandatory to treat the perturbations of the electromagnetic field necessary to study photon production, either real or virtual.
 
Here we follow \cite{D'Hoker:2009mm} to construct a family of back reacted geometries that include the black D3-brane solution, for $B=0$, and extends to backgrounds with arbitrarily large constant magnetic fields. Our family of solutions is accommodated in the general ansatz introduced in \cite{D'Hoker:2009mm} and has a number of nice features that make it an ideal toy model for addressing the effect of a background magnetic field on physical observables of interest. Such features include regularity of the fields on and outside the horizon and asymptotic AdS boundary conditions.

To find the photon production in the field theory side, we need to perturb the electromagnetic potential in the dual gravitational solution. The existence of a background magnetic field causes the effect of such perturbation on the stress-energy tensor to be of the same order as the perturbation itself, and therefore the perturbations of the metric and the electromagnetic potential have to be computed simultaneously.

Since the equations of motion for the metric and the vector potential are coupled, we need to use the method developed to compute correlators for mixing operators \cite{Kaminski:2009dh}.

We limit our study of the photon production to directions parallel and perpendicular to the background magnetic field and leave the general direction to be addressed in future work, where other interesting observables will be investigated. In the present paper we also compute the electric DC (i.e. zero-frequency) conductivity and compare it to the one of a plasma at the same temperature in the absence of the magnetic field.

We find that for strong enough background fields, the total production rate integrated over all propagation direction is larger than the isotropic rate independently of the frequency of the photons. Moreover, the stronger the magnetic field is the brighter the plasma glows compared to the case with no field. The conductivity drops to zero only in the direction perpendicular to the background field as soon we turn it on, while the conductivity in the direction of the field grows with it, which is consistent with a classic interpretation. These results indicate that the effect of a background magnetic field is distinguishable from that of other sources of anisotropy studied previously.

A real-world plasma undergoes a non-trivial time evolution, and in the current paper we assume staticity motivated by considering time scales much shorter than the characteristic evolution scale, which is as long as a few fm/c (see e.g. \cite{fmoverc}). We will consider the impact of the hydrodynamic evolution elsewhere.

The paper is organized as follows. In Sec.~\ref{sec1} expressions for the spectral densities and total production of photons are obtained when a background magnetic field is present. In Sec.~\ref{sec2} we construct the family of gravitational backgrounds that we will use. In Sec.~\ref{sec3} we compute the photon production rate and conductivity from holography and present our results. We analyze our results in Sec.~\ref{analysis} and conclude with a discussion and an outlook in Sec.~\ref{sec4}.


\section{Photon production in the presence of a strong magnetic field}
\label{sec1}

What we would like to do is to consider a gauge theory in the presence of a very intense magnetic field, that for definiteness we will point in the direction $z$. To be able to use the gauge/gravity correspondence, we will work in a four-dimensional ${\cal N}=4$ super Yang-Mills (SYM) with gauge group $SU(\nc)$, at large $\nc$ and large 't~Hooft coupling $\lambda=\gym^2\nc$.

The matter fields in the theory will be in the adjoint representation of the gauge group and have vanishing masses. With an abuse of language, we will refer to these matter fields as `quarks'. 

To study photon production we turn on a dynamical photon by including an $U(1)$ kinetic term in the SYM action and, following \cite{CaronHuot:2006te}, couple this photon to the conserved current associated with a $U(1)$ subgroup of the anomaly free global $SU(4)$ $R$-symmetry of the theory.

The action for the resulting $SU(\nc)\times U(1)$ theory is then
\be
S = S_{SU(\nc)} -\frac{1}{4}\int d^4x\left(  F^2 - 4e \, A^\mu J^\mt{EM}_\mu\right) \,,
\label{fulllagr}
\ee
where $F^2 = F_{\mu\nu}F^{\mu\nu}$ and $F_{\mu\nu} = \partial_\mu A_\nu - \partial_\nu A_\mu$ is the $U(1)$ gauge field strength, $e$ the electromagnetic coupling, and the electromagnetic current can be chosen to be
\be
J^\mt{EM}_\mu = \bar{\Psi} \gamma_\mu \Psi 
+ \frac{i}{2} \Phi^* \left( {\cal D}_\mu \Phi \right) 
- \frac{i}{2} \left( {\cal D}_\mu \Phi \right)^* \Phi \,,
\label{current}
\ee 
with an implicit sum over the flavor indices and which is invariant under the generator $t^3\equiv$diag$(\frac{1}{2},\frac{-1}{2},0,0)$ of the $R$-symmetry.

The matter fields are differentiated using the full $SU(\nc)\times U(1)$ connection, ${\cal D}_\mu=D_\mu-ieA_\mu$.

\begin{figure}
\begin{tabular}{cc}
\includegraphics[width=0.53 \textwidth]{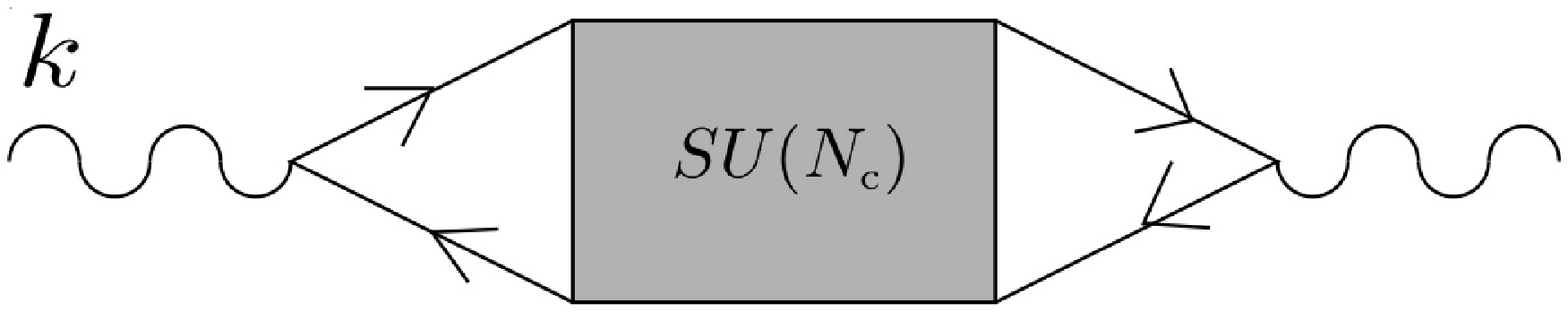}
 &
\includegraphics[width=0.4 \textwidth]{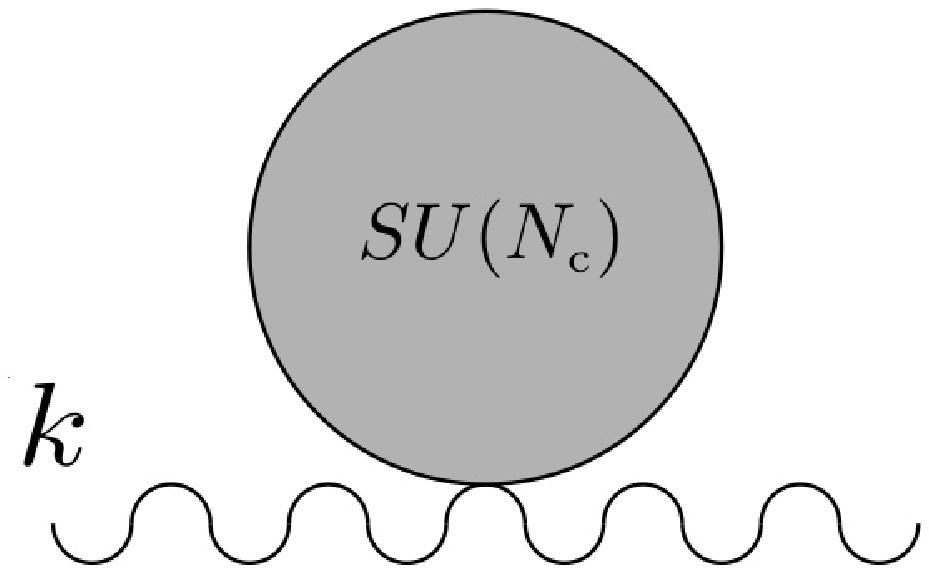}
\\
\end{tabular}
\caption{Diagrams contributing to the two-point function of electromagnetic currents to leading order in the electromagnetic coupling $\alpha_\mt{EM}$. The external, wavy lines correspond to photons with momentum $k$, while the shaded blobs denote full resummations of the $SU(\nc)$ diagrams, to all orders in $\lambda$. We see here that no dynamical photons are involved}\label{diag}
\end{figure}

At a first glance it seams as we should use the gravitational dual of the full $SU(\nc)\times U(1)$ theory, but as argued in \cite{CaronHuot:2006te}, our calculation can be done in the  $SU(\nc)$ theory, of which the dual in the presence of a strong magnetic field was studied in \cite{D'Hoker:2009mm}. The reason for this is that, given the smallness of the electromagnetic coupling, the results for the two point function can be kept to leading order in $\alpha_\mt{EM}$, and fully non-perturbative in the 't~Hooft coupling $\lambda$ of the $SU(\nc)$ theory, taking full advantage of the holographic computation without introducing a dynamical photon. In Fig. \eqref{diag} this situation is described diagrammatically, the shaded blobs indicate the all-order resummations of the $SU(\nc)$ theory diagrams and the external legs represent the $U(1)$ non-dynamical photons. 

The differential photon production is given by \cite{lebellac,CaronHuot:2006te,Mateos:2007yp}
\bea
\frac{d\Gamma}{d\vec k} = \frac{e^2}{(2\pi)^3 2|\vec k|}\Phi(k)\sum_{s=1,2} \epsilon^\mu_{(s)}(\vec k)\,  \epsilon^\nu_{(s)}(\vec k)\, \chi_{\mu\nu}(k)\Big|_{k^0=|\vec k|}\,,
\label{diff}
\eea
where $k^\mu=(k^0,\vec k)$ is the null wave vector for the photon and $\Phi(k)$ is the distribution function on the photon momentum. In our case we can consider that the QGP is in thermal equilibrium and we must use the Bose-Einstein distribution $n_B(k^0)=1/(e^{k^0/T}-1)$. The spectral density is given by $\chi_{\mu\nu}(k)=-2 \mbox{ Im } G^\mt{R}_{\mu\nu}(k)$, where the retarded correlator of the two electromagnetic currents $\jem_\mu$ is 
\bea
G^\mt{R}_{\mu\nu}(k) = -i \int d^4x  \, e^{-i k\cdot x}\, \Theta(t) \vev{[\jem_\mu(x),\jem_\nu(0)]}.
\eea

\begin{figure}
    \begin{center}
        \includegraphics[width=0.40\textwidth]{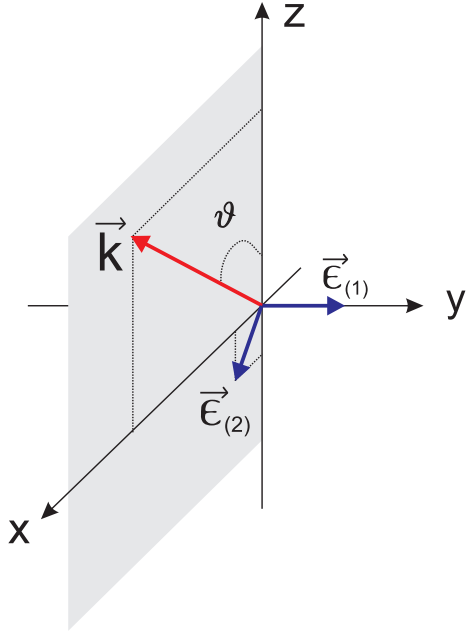}
        \caption{Photon momentum and polarization vectors. Because of the rotational symmetry in the $xy$-plane, the momentum can be chosen to be contained in the $xz$-plane, forming an angle $\vartheta$ with the $z$-direction. $\vec \epsilon_{(1)}$ is oriented along the $y$-direction and $\vec\epsilon_{(2)}$ is contained in the $xz$-plane, orthogonally to $\vec k$.}
        \label{momentum}
    \end{center}
\end{figure}

The two terms of the sum in (\ref{diff}) account for the number of photons emitted with polarization vector $\vec\epsilon_{(s)}$. The two polarization vectors and $\vec k$ are orthogonal to each other. We can use the unbroken rotational $SO(2)$ symmetry in the $xy$-plane to, without  loss of generality, point $\vec k$ to be contained in the $xz$-plane and denote its direction by the angle $\vartheta$ that it sustains with respect to the $z$-axis -- see Fig.~\eqref{momentum}.

As will be seen in the following section, the direct calculation in the dual gravitational theory (\cite{D'Hoker:2009mm}) results in a metric that asymptotically approaches
\be
{ds^2}_{r\rightarrow \infty}=r^2[-dt^2+ C_1 (dx^2+dy^2)+C_2 dz^2]+\frac{dr^2}{ r^2}, \label{gTC5}
\ee
where $C_1, C_2$ are constant factors, providing the field theory with the flat line element
\be
{ds^2}_{FT}=dt^2+ C_1 (dx^2+dy^2)+C_2 dz^2, \label{gTC}
\ee
that trivially corresponds to the Minkowski metric after scaling the boundary coordinates.

Even though such scaling is of no consequence, we need to point it out here since it will, in practice, turn out to be easier to keep the factors correct if we write the expressions in the field theory side using the metric (\ref{gTC}), even if that means that we will have to keep track of the nature of the indices, covariant or contravariant, in expressions where normally it would make no difference. We need to be emphatic that this is just a way to keep track of some factor, since, as it should, all the final expressions will be invariant, and in the final results the factors cancel appropriately.

For instance, for the momentum of the photon to be null, we set its covariant components to be
\be
k_\mu = k_0 (1,\sqrt{C_1}\sin \vartheta, 0, \sqrt{C_2}\cos \vartheta) \,.
\ee
And similarly for the polarization vectors to be orthogonal and chosen unitary their contravariant components can be taken to be
\be
{\epsilon_{(1)}}^\mu=(0,0,\frac{1}{\sqrt{C_1}},0)\sac {\epsilon_{(2)}}^\mu = (0,\frac{1}{\sqrt{C_1}}\cos \vartheta, 0, \frac{-1}{\sqrt{C_2}} \sin \vartheta) \,.
\ee
For photons with polarization $\vec\epsilon_{(1)}$ the production is proportional to $\epsilon_{(1)}^\mu\epsilon_{(1)}^\nu \chi_{\mu\nu}=\tfrac{\chi_{yy}}{C_1} \sim \tfrac{ \mbox{Im}\, \langle \jem_y \jem_y \rangle}{C_1}$, while for those with polarization $\vec\epsilon_{(2)}$ it is proportional to\footnote{Note that $\chi_{xz}=\chi_{zx}$, see e.g.~\cite{CaronHuot:2006te}}.
\be
\epsilon^\mu_{(2)}\,  \epsilon^\nu_{(2)}\, \chi_{\mu\nu} = (
\frac{1}{C_1}\cos^2 \vartheta \, \chi_{xx} + \frac{1}{C_2}\sin^2 \vartheta \, \chi_{zz} 
- \frac{2}{\sqrt{C_1 C_2}} \cos \vartheta \sin \vartheta  \, \chi_{xz} )\,. 
\label{combination}
\ee
In what follows, we will specialize to the cases $\vartheta=0$ and $\pi/2$ so, in view of the expressions in this section, we see that what we need to compute are the Green functions $G^\mt{R}_{xx} $ and $G^\mt{R}_{yy} $ for photons propagating in the $z$ direction, and $G^\mt{R}_{yy} $ along with $G^\mt{R}_{zz} $ for photons propagating in the $x$ direction. Once we have obtained these quantities, we can add them up according to the expressions above. In the following section we will see how these correlators can be obtained from the dual gravitational setup.


\section{Dual gravitationa setup}
\label{sec2}

The dual gravitational background for the theory described in section (\ref{sec1}) was studied in \cite{D'Hoker:2009mm}. This is a numerical solution that interpolates between the product of a BTZ black hole with $T^2$ near the horizon, and $\mathrm{AdS}_5$, up to the scaling mentioned above, close to the boundary.

The action to be considered is the one of a five-dimensional Einstein-Maxwell theory with a negative cosmological constant
\be
S=-\frac{1}{16 \pi G_5}\int d^5x\sqrt{-g}(R+F^{mn}F_{mn}-\frac{12}{L^2})+S_{\mathrm{bdry}}, \label{graction}
\ee
where the only boundary term relevant for our calculations will be the one derived in section (\ref{sec3}), as normally in the correspondence $G_5=\frac{\pi}{2 \nc^2}$, and for convenience, $L$ will be set to unity. It is important to clarify that the field strength $F$ that appears here is not the same as the one in (\ref{fulllagr}), since the latter is the one living in the field theory while the one in (\ref{graction}) is a field in the gravity side which perturbations $A_\mu$ will be seen below to be dual to the current $\jem_\mu$ in the field theory. Since the vector potential of the gauge theory in (\ref{fulllagr}) will not appear anywhere below, the usage of the same letter for the latter and the perturbations $A_\mu$ of the U(1) field in the gravity side should not generate confusion.

The general ansatz used in \cite{D'Hoker:2009mm} to asymptotically approach $\mathrm{AdS}_5$ with a magnetic field tangent to the boundary directions can be rewritten as
\be
ds^2=-U(r) dt^2+\frac{1}{U(r)}dr^2+V(r)(dx^2+dy^2)+W(r)dz^2, \label{bgm}
\ee
for the metric and
\be
F_{BG}=B\ dx\wedge dy, \label{bgf}
\ee
with constan $B$ for the magnetic field strength.

The Maxwell equations are automatically satisfied and the Einstein equations coming from the variation of (\ref{graction}) reduce to
\begin{align}\label{eqfond}\nonumber
 2 W(r)^2 \,\bigg[4 B^2  +V(r)\, \Big(U'(r)\, V'(r)+U(r) \, V''(r)\Big)\bigg]\, -\, V(r)\, W(r)\,\, \bigg[2 V(r) \\ \nonumber
 \times\Big(U'(r) W'(r)+U(r) W''(r)\Big)+U(r) V'(r) W'(r)\bigg]+U(r) V(r)^2 W'(r)^2 &= 0,\\
  4 V(r) W(r)^2 V''(r)-2 W(r)^2 V'(r)^2-V(r)^2\Big(W'(r)^2-2 W(r) W''(r)\Big) &= 0,  \\ \nonumber
  W(r) \bigg[-8 B^2+6 V(r)^2 \Big(U''(r)-8\Big)+6 V(r) U'(r) V'(r)\bigg]+3 V(r)^2 U'(r) W'(r) &= 0,\\  \nonumber
 W(r) \Big(4 B^2+2 V(r) U'(r) V'(r)+U(r) V'(r)^2 -24 V(r)^2\Big)+V(r) W'(r)\\ \nonumber
\times  \Big(V(r) U'(r)+2 U(r) V'(r)\Big) &= 0,
\end{align}

\noindent where the fourth is a constrain equation for the initial data, and once it is satisfied at some radius, the evolution provided by the other three equations will ensure that it remains satisfied.

There are two analytic solutions to (\ref{eqfond}) that will be of interest here. One of them is given by
\be
U_{BTZ}(r)=3(r^2-{r_h}^2),\,\,\,V_{BTZ}(r)=\frac{B}{\sqrt{3}} \,\,\,\mathrm{and} \,\,\, W_{BTZ}(r)=3r^2, \label{BTZ}
\ee
and represents the aforementioned BTZ black hole with horizon at $r_h$ times a two dimensional space that in \cite{D'Hoker:2009mm} is taken to be $T^2$, where they successfully constructed a numerical solution that interpolates between (\ref{BTZ}) for $r\rightarrow r_h$ and (\ref{gTC5}) for $r\rightarrow \infty$.

The other solution we want to mention is given by
\bea
U_{BB}(r)&=&\left(r+\frac{r_h}{2}\right)^2\left(1-\frac{\left(\frac{3}{2}r_h\right)^4}{\left(r+\frac{r_h}{2}\right)^4}\right), \nonumber \\
V_{BB}(r)&=&\frac{4 V_0}{9 r_h^2}\left(r+\frac{r_h}{2}\right)^2, \label{BBr}\\
W_{BB}(r)&=&\frac{4}{3} \left(r+\frac{r_h}{2}\right)^2, \nonumber
\eea
which is the black D3 brane solution with the radial coordinate $\tilde{r}$ shifted to $\tilde{r}=r+\frac{r_h}{2}$, consequently $\tilde{r}_{h}=\frac{3}{2}r_h$, and some constant scaling of the $x,y$ and $z$ coordinates.

The reason to shift the radial coordinate and do the scaling in the black brane solution is that in this way, close to the horizon (\ref{BTZ}) and (\ref{BBr}) coincide in their metric function $U(r)$ up to first order in powers of $(r-r_h)$ and in $W(r)$ up to leading order. This feature will prove useful below.

To find the family of solutions that we are interested in, we notice that the analysis of the system of equations (\ref{eqfond}) reveals that the space of solutions is four dimensional, and therefore if a series expansion was to be performed, it would only admit four independent coefficients.

In particular, the solution (\ref{BTZ}) can be written exactly as a power series on $(r-rh)$ given by
\bea
U_{BTZ}(r)&=&6 rh (r-r_h)+3(r-r_h)^2, \nonumber \\
V_{BTZ}(r)&=&\frac{ B }{\sqrt{3}}, \label{BTZhor}\\
W_{BTZ}(r)&=&3 {r_h}^2+6 r_h (r-r_h)+3(r-r_h)^2, \nonumber
\eea
while the solution (\ref{BBr}) can be written as the series
\bea
U_{BB}(r)&=&6 r_h (r-r_h)-2 (r-r_h)^2+\frac{8 }{3 r_h}(r-r_h)^3+{\cal O}(4) \nonumber \\
V_{BB}(r)&=&V_0+\frac{4 V_0 }{3 r_h}(r-r_h)+\frac{4 V_0 }{9 r_h^2}(r-r_h)^2, \label{BBP}\\
W_{BB}(r)&=&3 r_h^2+4 r_h (r-r_h)+\frac{4}{3} (r-r_h)^2, \nonumber
\eea
where ${\cal O}(4)$ indicates terms of fourth order or higher and the other two series terminate as shown.

Observing (\ref{BTZhor}) and (\ref{BBP}), the way in which we will look for solutions  will be to permit (\ref{BTZhor}) to be modified  according to
\bea
U_P(r)&=&U_{BTZ}+\sum _{\text{i}=2}^\infty U_i(r-r_h)^i, \nonumber \\
V_P(r)&=&V_{BTZ}+\sum _{\text{i}=0}^\infty V_i(r-r_h)^i, \label{BTZP}\\
W_P(r)&=&W_{BTZ}+\sum _{\text{i}=1}^\infty W_i(r-r_h)^i, \nonumber
\eea
keeping the first two coefficients in the series for $U$ (considering the zero order) and the first in the series for $W$ identical to those in both (\ref{BTZhor}) and (\ref{BBP}). Doing so we specify three out of the four available parameters, leaving only one free coefficient for the rest of the expansion, and it can consistently be taken to be $V_0$.

In this way we have implemented a one parameter, $V_0$, perturbation of (\ref{BTZ}) near the horizon that will allow us to construct a family of numerical solutions which elements smoothly approach the black brane solution (\ref{BBr}) when $B\rightarrow 0$ as long as $V_0$ does not vanish and, as they should, for non vanishing $B$ the exact solution (\ref{BTZ}) is recovered when $V_0$ is set to zero. All intermediate cases provide interesting solutions, but the ones satisfying $\frac{B}{V_0}\gg 1$ will correctly interpolate between (\ref{BTZ}) for $r\rightarrow r_h$ and (\ref{gTC5}) for $r\rightarrow \infty$ as desired. A relevant fact about this family of solutions is that all its elements have the same temperature $T=\frac{\tilde{r}_{h}}{\pi}=\frac{3r_h}{2\pi}$ as the one associated with (\ref{BBr}) by the regularity of the Euclidean continuation.

This construction will give us the possibility of checking that the results we obtain for the photon production reduce continuously to the analytic ones found in \cite{CaronHuot:2006te,Mateos:2007yp} when $B=0$.

To solve (\ref{eqfond}) perturbatively around $r_h$, we can insert (\ref{BTZP}) in (\ref{eqfond}) and solve for all the coefficients in terms of $V_0$, procedure which for most cases will render an infinite sum that will not be exactly sumable, and the general analytic solution to (\ref{eqfond}) is not known.

So in practice, and given the singular character of (\ref{eqfond}) at $r=r_h$, we solve for the first few coefficients and use them to give initial data for the solution arbitrarily close to $r_h$ to integrate numerically towards $r\rightarrow\infty$. Following these steps, we have been able to carry the numerical integration for arbitrarily high values of $\frac{B}{V_0}$ maintaining high precision at the expense of longer computing times.

\begin{figure}
\begin{center}
\begin{tabular}{cc}
\setlength{\unitlength}{1cm}
\hspace{-0.9cm}
\includegraphics[width=7cm]{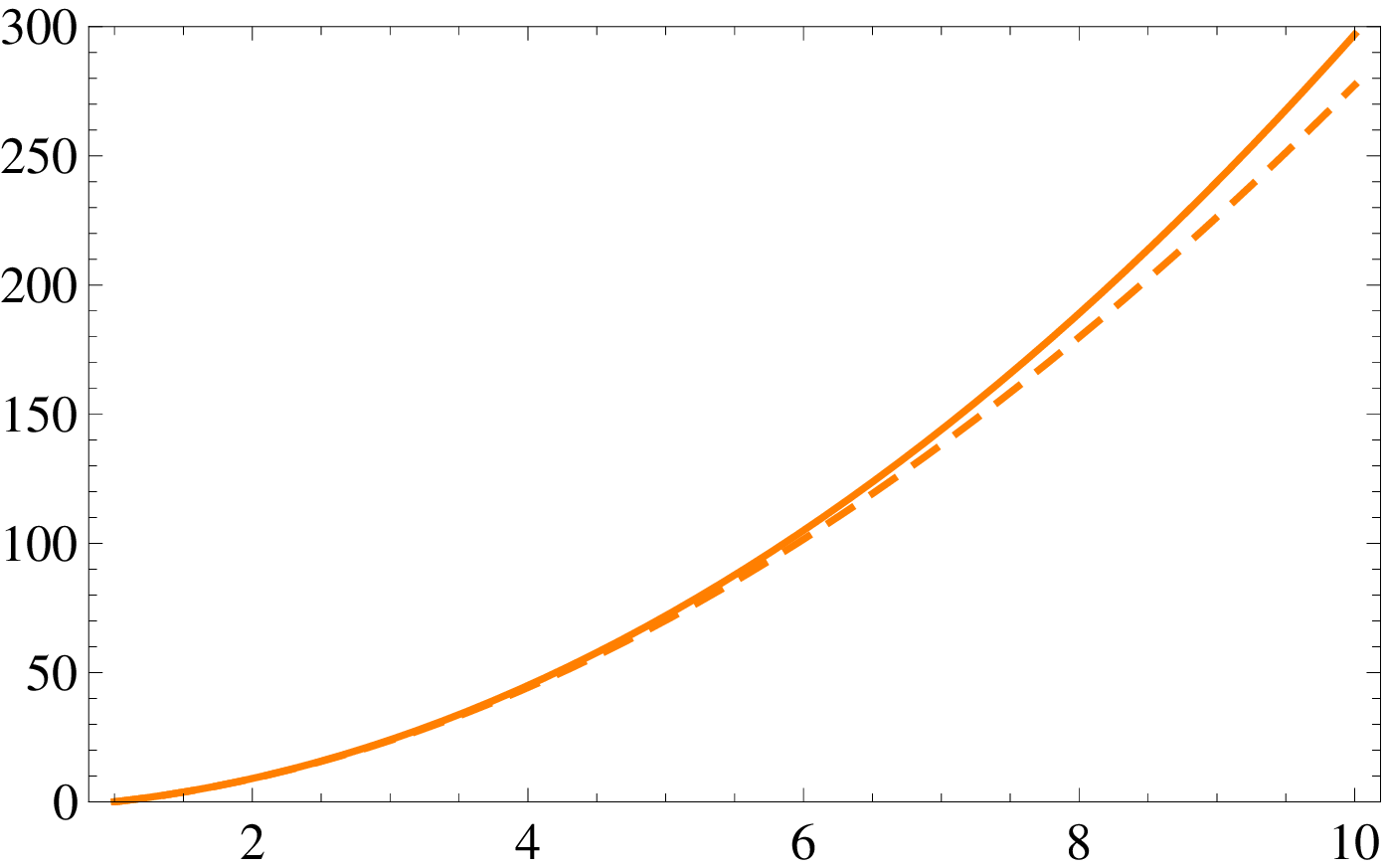} 
\qquad\qquad & 
\includegraphics[width=7cm]{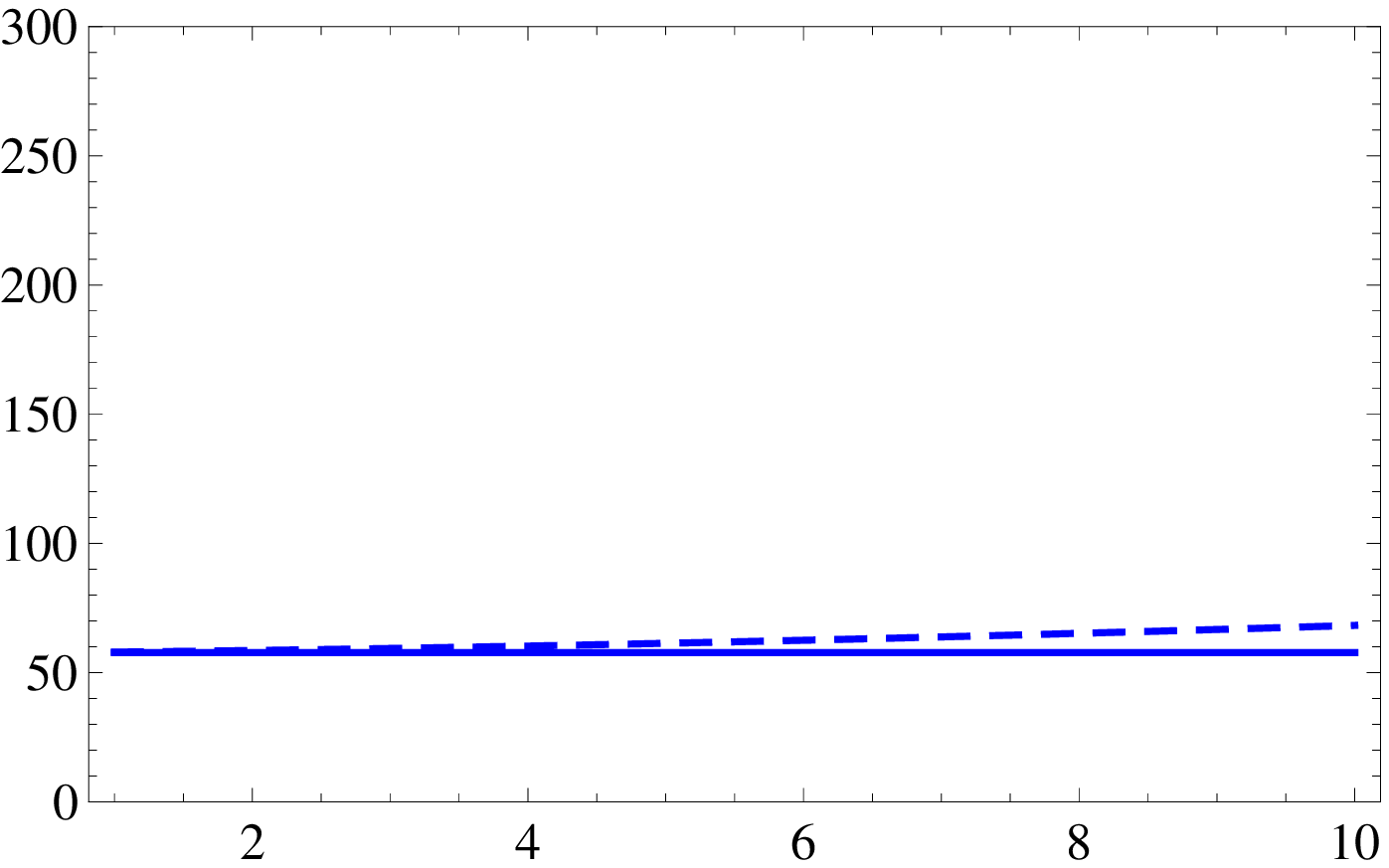}
\qquad
  \put(-455,110){$U$}
         \put(-250,-10){$r/rh$}
         \put(-215,110){$V$}
         \put(-17,-10){$r/rh$}
\\
(a) & (b)\\
& \\
\hspace{-0.9cm}
\includegraphics[width=7cm]{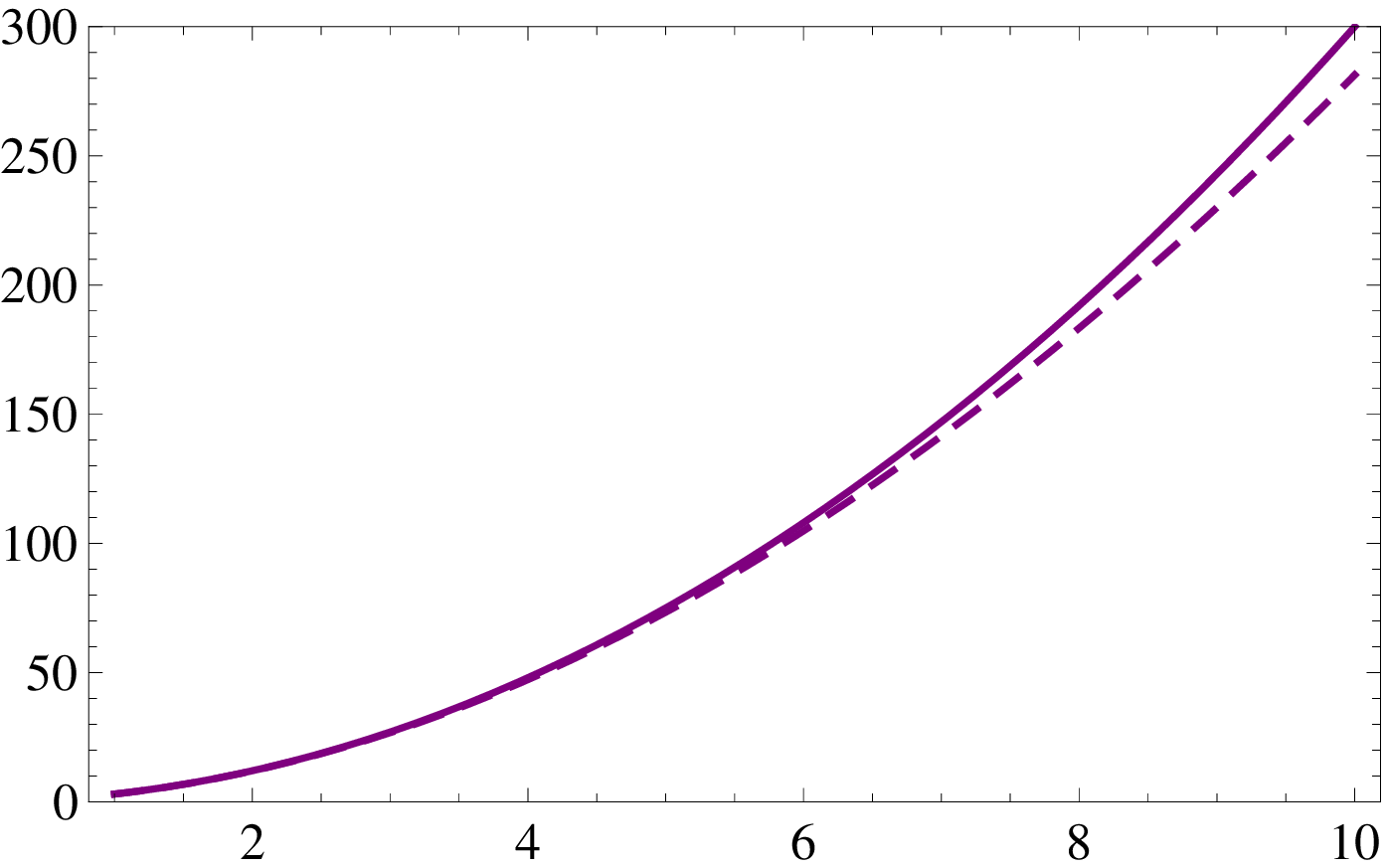} 
\qquad\qquad & 
\includegraphics[width=7cm]{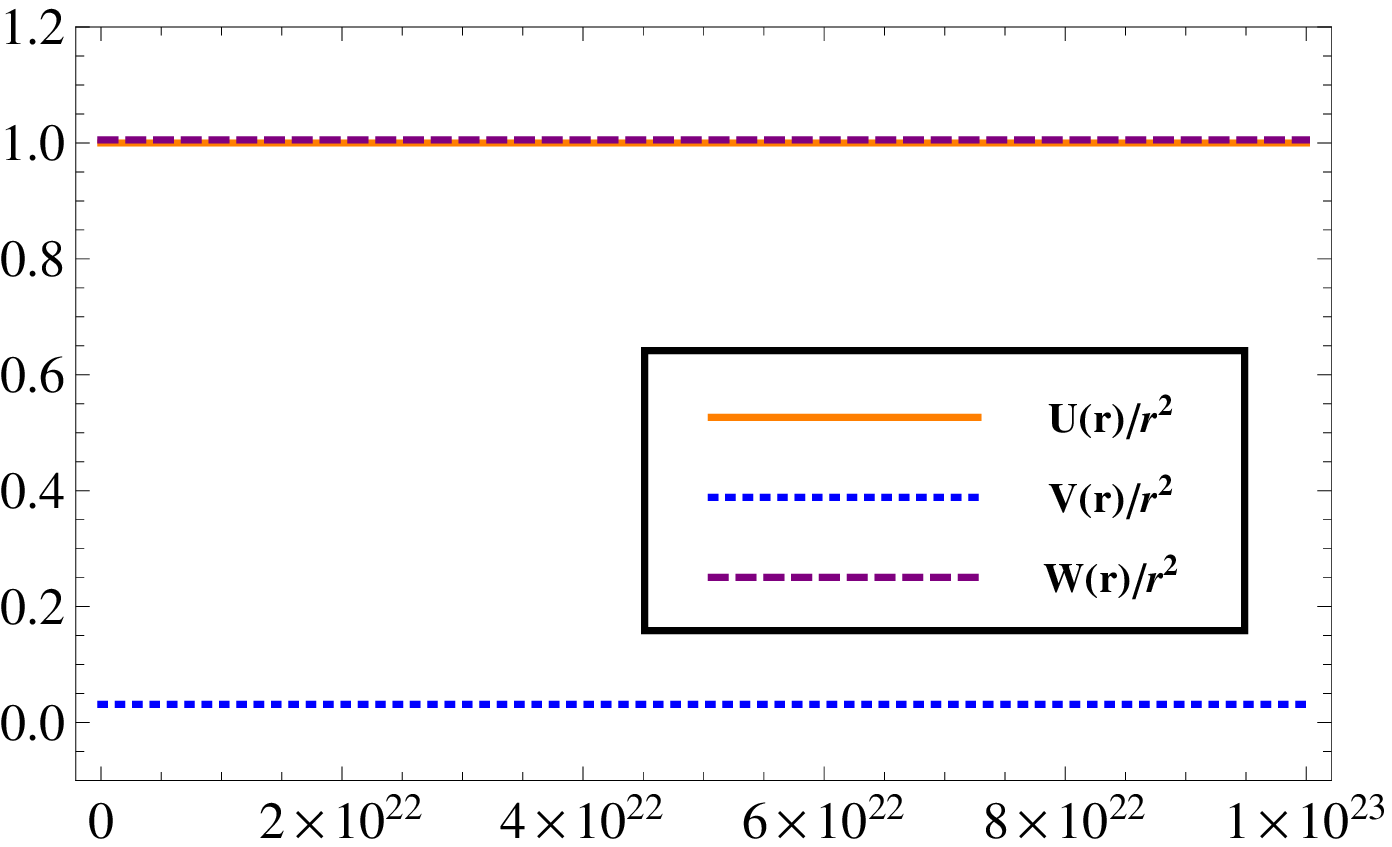}
\qquad
 \put(-455,110){$W$}
         \put(-250,-10){$r/rh$}
         \put(-215,40){\rotatebox{90}{$U\quad ,V\quad ,W$}}
         \put(-17,-10){$r/rh$}
         \\
(c)& (d) 
\end{tabular}
\end{center}
\caption{\small In dashed lines, the plots of a representative numerical solutions to \eqref{eqfond}, compared to the BTZ solution, in solid lines, near the horizon in (a) for $U$, (b) for $V$ and (c) for $W$, and compared with the AdS${}_{5}$ solution at large $r$ in (d).}
\label{plotf}
\end{figure}

A generic solution with $\frac{B}{V_0}\gg 1$ is shown close to $r_h$ in Fig. \eqref{plotf} a,b and c where the closeness to the BTZ solution can be appreciated, and for large radius in Fig. \eqref{plotf} d, where the asymptotic approach to $\mathrm{AdS}_5$ can be seen.

Now that we have a family of gravitational backgrounds with a constant magnetic field, which are dual to the field theory described in section (\ref{sec1}), we are in position to compute the photon production for different field intensities of the background field $B$ in such theories.

As we mentioned before, and originally claimed in \cite{CaronHuot:2006te}, in order to obtain the photon production to leading order in $e$ one can compute the correlators which enter in \eqref{diff} from the $SU(\nc)$ gauge theory with no contribution from dynamical photons whatsoever. These correlators can be computed holographically for strong 't Hooft coupling in the large $\nc$ limit. For this, we must identify the gauge field on the gravity setup which is dual to the conserved current of interest in the gauge theory.

In this case we need to study the behavior of perturbations around the background provided by (\ref{bgm}) and (\ref{bgf}). In particular, to study photon production, we need to examine the behavior of the perturbations $(A_\mu, A_r)$ $(\mu=0, \ldots, 3)$ of the U(1) gauge field that appears in (\ref{graction}) and provides the dual of the conserved current that we are interested in. A crucial point of the calculation is that given the presence of a finite background magnetic field, the contribution to the stress-energy tensor of the electromagnetic perturbations is of the same order as themselves, making it necessary to introduce perturbations of the background metric as well \footnote{This is a crucial point in this work and makes a substantial difference with related previous works and in particular with our own previous version of the present work. In section \ref{analysis} we analyze the importance and consequences of introducing such perturbations.}. 

To keep track of the order of all contributions, we write 
\be 
g_{\mu\nu}={g^{BG}}_{\mu\nu} + \epsilon\ h_{\mu\nu} \qquad {\rm and} \qquad F=F^{BG}+\epsilon\ dA, \label{fildpert}
\ee
where $BG$ refers to background fields (\ref{bgm}) and (\ref{bgf}), while $\epsilon$ is the perturbation parameter.

To perturb the action around the minimum provided by (\ref{bgm}) and (\ref{bgf}), we insert (\ref{fildpert}) in to (\ref{graction}) and solve the equations of motion coming from the variation with respect to both $h$ and $A$ when only terms up to first order in $\epsilon$ are kept.

Working in the gauge $A_r=0$ leaves the gauge field $A_\mu$ to be the desired dual to the conserved electromagnetic current $\jem_\mu$ of the gauge theory, and all the equations of motion decouple in two sets.  One of the sets involves the components $t, x$ and $z$ of the $A$ field along with the components $ty, xy, yz$ and $yr$ of the metric perturbation, while the other set of equations involves only $A_y$ and all the remaining components of $h$. These general equations are lengthy and not particularly enlightening, so we present them in the appendix and only the relevant particular cases will be part of the main text below.

According to the prescription of \cite{duality2,duality3}, correlation functions of $\jem_\mu$ can be calculated by varying the gravity partition function with respect to the value of $A_\mu$ at the boundary of the spacetime \ref{bgm}. To follow this procedure, and given that\footnote{To zero order in $\epsilon$.} the boundary is perpendicular to $r$, we further wish to impose the gauge $h_{\mu r}=0$ for the metric perturbations \footnote{Specifically the reason to choose this gauge is that, as will be seen below, it will be necessary to find solutions to the coupled field equations in which only one component of the gauge field is different form zero at the boundary and all the components of the metric perturbation vanish. It can be seen that the only way to make the components $h_{\mu r}$ vanish at the boundary for a non trivial solution is to gauge them away from the beginning. This statement becomes apparent in both sides of the duality. In the gravity side it expresses itself in the fact that the field equations permit to solve for $h_{\mu r}$ algebraically in terms of the rest of the fields, so it cannot be forced to vanish at the boundary for the particular solution with only one component of the gauge field with a value different form zero at the boundary. This is consistent with the fact that this components of the metric perturbation can not be taken as independent sources in the gauge theory, and, form the perspective of the four dimensional theory, cannot be eliminated by a gauge transformation.}. After fixing this gauge we are left with twenty equations and fourteen fields in total, so at first sight it could seam like the system was overdetermined, but since this is only due to a gauge choice, not all the equations will be independent and the system will be well posed. In all the following cases we have explicitly checked that the solutions of a set of independent equations also solves all the remaining ones.

To solve the field equations in the gravity side it will be useful to notice that, even if the theory we are considering is anisotropic due to the presence of the background magnetic field $F_{BG}$, it is translationally invariant along the gauge theory directions, so we can Fourier decompose all the fields as
\be
\Phi(x^\mu, r) = \int \frac{d^4k}{(2\pi)^{4}} \, 
e^{ix^\mu k_\mu} \, \Phi(k_\mu, r) \,,\qquad k_\mu = k_0 (1,\sqrt{C_1}\sin \vartheta, 0, \sqrt{C_2}\cos \vartheta)\, ,
\label{fourier}
\ee
where the constants appearing in the components $k_\mu$ where discussed in section (\ref{sec1}) and $\Phi$ stands for any of the components of the fields $A$ and $h$.

To compute the total photon production it will be necessary to take the trace of the spectral function. To the perturbative order we are interested in this can be consistently done using the background metric alone.


\section{Photon production computed holographycally}
\label{sec3}

The method to compute the correlation functions for the electromagnetic current via the Lorentzian AdS/CFT correspondence is given in \cite{recipe} (see also \cite{PSS,PSS2,KS}). To follow this prescription we must find in the action \eqref{graction} those terms with second order derivatives with respect to the radial direction, integrate by parts to obtain the boundary term and evaluate this last on shell. The second variation of the resulting boundary action with respect to the value of the gauge field at the boundary $A_{\mu}^{bdry}$, which play the role of a source for the currents in the field theory, gives the desired correlation functions. The way to take the variation of the boundary action with respect to the values of the fields on it when the bulk field equations are coupled was investigated in \cite{Kaminski:2009dh}, so we will resort to this method as our equations for $A$ and $h$ can not be separated.

The part of the boundary action with second order terms in the perturbation fields, $A$ and $h$, is
\begin{multline}  \label{GBA}
S_{\frac{1}{\epsilon}} = - \frac{1}{8\pi G_5} \int d^4x\ \bigg\{ V(r)\sqrt{W(r)}U(r) \Big[-U^{-1}(r)\ A_t A'_t \\ 
+ V^{-1}(r)\ A_x A'_x + V^{-1}(r)\ A_y A'_y + W^{-1}(r)\ A_z A'_z \Big] \bigg\}_{r=\frac{1}{\epsilon}} \\
 - \frac{1}{16\pi G_5} \int d^4x\ \bigg\{ V(r)\sqrt{W(r)}\  \Big[ \mathcal{O}(h^2) + \mathcal{O}(hh')\Big] \bigg\}_{r=\frac{1}{\epsilon}},
\end{multline}
where taking the limit $\epsilon\rightarrow 0$ is understood and zeroth, first and higher than second order terms in the perturbation fields are not written. We will see further ahead that the $\mathcal{O}(h^2)$ and $\mathcal{O}(hh')$ terms in the boundary action do not contribute to the retarded Green function as we are taking second variations with respect to $A^{bdry}$. The only terms which could have contributed are those of order $\mathcal{O}(h'^2)$, $\mathcal{O}(Ah')$ and $\mathcal{O}(A'h')$, but none of these appear in the boundary action. 
To evaluate (\ref{GBA}) on shell we will solve the field equations in the gravity side in a number of situations studied below. As a warm up, and given that our family of geometries permits it, we will start by carrying out the calculation for the $B=0$ case, where no coupling of the field equations occur.
 

\subsection{$B=0$ limit}

Given that our family of solutions lets us take the limit $B=0$ smoothly, it will be useful to compute the spectral densities for photon production for this case, which is analytic and corresponds to the one found in \cite{CaronHuot:2006te,Mateos:2007yp}.

In the absence of a background field $B$, the contribution of the perturbation $A$ to the stress energy tensor vanishes to first order, so we can consistently turn off the metric perturbations $h$.

The boundary action (\ref{GBA}) simplifies to 
\begin{multline}\label{boundaA}
S_\frac{1}{\epsilon}=
-\frac{1}{8\pi G_5}\int d^4x\, \left[V(r)\sqrt{W(r)}U(r) \Big( -U^{-1}(r)A_tA_t'+V^{-1}(r)A_xA_x'\right.\\
\left.+V^{-1}(r)A_yA_y'+W^{-1}(r) A_zA_z'\Big)\right]_{r=\frac{1}{\epsilon}}\, , 
\end{multline}
where the limit $\epsilon\rightarrow 0$ is intended and only Maxwell's equations have to be solved.

We may also set $\vartheta=0$, since in the absence of a background magnetic field there is an $SO(3)$ symmetry that permits to align the photon momentum along a particular direction. In this case we only need to compute $G^\mt{R}_{yy}$ and $G^\mt{R}_{xx}$, while the equation to solve for either $A_x$ or $A_y$ is 
\bea
\left(\sqrt{-g} g^{rr}g^{ii}A'_i\right)'-\sqrt{-g} g^{ii}\left(g^{tt}k_t^2+g^{zz}k_z^2\right)A_i=0\,,
\label{eomy}
\eea
where the index $i$ is either $x$ or $y$.

From the boundary action \eqref{boundaA} and the equation (\ref{eomy}) we see that we can carry on the calculation for the spectral density $\chi_{ii}$ independently of all the others and the calculation is very similar to the one in \cite{Mateos:2007yp}. 

By plugging (\ref{BBr}) in (\ref{eomy}) we explicitly obtain the equation of motion for either field given by 
\begin{multline}
\left[\left(r+\frac{r_h}{2}\right)^3\left(1-\frac{\left(\frac{3}{2}r_h\right)^4}{\left(r+\frac{r_h}{2}\right)^4}\right)A'_{B=0}(\wn, r)\right]'\\
+\frac{{k_0}^2}{\left(r+\frac{r_h}{2}\right)} \left[{\footnotesize{\left(1-\frac{\left(\frac{3}{2}r_h\right)^4}{\left(r+\frac{r_h}{2}\right)^4}\right)^{-1}-1}}\right]A_{B=0}(\wn, r)=0,\label{eqb0}
\end{multline}
where in writing the components of the photon momentum it was taken in to account that the solution (\ref{BBr}) leads to a metric in the field theory given by
\be
ds^2=-dt^2+\frac{4 V_0}{9 r_h}(dx^2+dy^2)+\frac{4}{3}dz^2. \label{partmet}
\ee

The solution to this equation that is infalling at the horizon reads
\begin{multline}
A_{B=0}=\left(1-\frac{9 {r_h}^2}{(2 r+{r_h})^2}\right)^{-\frac{i \wn}{2}} \left(\frac{9 {r_h}^2}{(2 r+{r_h})^2}+1\right)^{\frac{\wn}{2}} \,\\
\times {}_2F_1\left(\frac{\left(1-i \right) \wn}{2}+1, \frac{\left(1-i\right) \wn}{2};1-i\wn;\frac{2 \left(r^2+{r_h} r-2 {r_h}^2\right)}{(2r+{r_h})^2}\right),
\end{multline}
where ${}_2F_1$ is an hypergeometric function and by using the temperature $T=\frac{3r_h}{2\pi}$ associated with our family of solutions, the dimensionless frequency
\be
\wn=\frac{k_0}{2\pi T}=\frac{k_0}{3 r_h},
\ee
has been introduced, as customarily in the quasi-normal mode literature. 

For the chosen angle, the sum over polarizations in (\ref{diff}) is equal to
\be
\chi_{B=0}=-2\, \mbox{Im}\,\frac{4 V_0}{9 {r_h}^2}\left({G^\mt{R}}^{yy}+{G^\mt{R}}^{xx}\right)_{B=0}=-4\, \mbox{Im}\,\frac{4 V_0}{9 {r_h}^2} G^\mt{R}_{B=0}\,,
\label{spectralisotropic}
\ee
and this last can be computed from the prescription in \cite{recipe} using
\begin{multline}\label{gyyb0}
G_{B=0}^\mt{R} = -\frac{1}{8\pi G_5 |A_{B=0}(\wn, \infty)|^2}   \\
\times \lim_{r \rightarrow \infty}  
 \frac{2}{\sqrt{3}}\left(r+\frac{r_h}{2}\right)^3\left(1-\frac{\left(\frac{3}{2}r_h \right)^4}{\left(r+\frac{r_h}{2}\right)^4}\right) A_{B=0}^*(\wn,r) A'_{B=0}(\wn, r) \, ,
\end{multline}
where 
\begin{multline}
A_{B=0}(\wn, \infty)=  \left(1-\frac{9 {r_h}^2}{(2 r+{r_h})^2}\right)^{-\frac{i \wn}{2}} \left(\frac{9 {r_h}^2}{(2 r+{r_h})^2}+1\right)^{\frac{\wn}{2}} \,\\
 \times\, {}_2F_1\left(\frac{\left(1-i \right) \wn}{2}+1,\frac{\left(1-i\right) \wn}{2}; 1-i\wn;\frac{1}{2}\right), \label{AB0inf}
\end{multline}
is the value that $A_{B=0}$ takes at $r\rightarrow\infty$.

The fact that the indices in (\ref{spectralisotropic}) are contravariant is understood given that the way (\ref{gyyb0}) is obtained from the gauge gravity correspondence, is by varying the boundary action twice with respect to the value of the field at the boundary ${A_y}^{bdry}$, which index is covariant, so, for the indices to be properly contracted, the result of this variation most have two contravariant indices. The same remark will apply to the variations with respect to the boundary values of the other fields necessary to compute the correlators in the remaining of the calculations.

Multiplying eq. (\ref{eqb0}) by $A_{B=0}^*(\wn, r)$ we see that the imaginary part of the expression inside the limit in (\ref{gyyb0}) turns out to be independent of $r$. It is actually easier to evaluate the limit at the horizon rather than at the boundary, and the result is
\be
\lim_{r \rightarrow r_h}  
 \frac{2}{\sqrt{3}}\left(r+\frac{r_h}{2}\right)^3\left(1-\frac{(\frac{3}{2}r_h)^4}{(r+\frac{r_h}{2})^4}\right)A_{B=0}^*(\wn,r) A'_{B=0}(\wn, r)=-i\frac{\wn 2^\wn (2\pi T)^2}{\sqrt{3}}\,.
\ee
One last thing to notice is that here we are computing densities with respect to the metric (\ref{gTC}), or \eqref{partmet} in this particular case, in the field theory, so to take into account the scaling necessary to turn the metric into the Minkowsky line element, our results have to be divided by $\sqrt{-h}$, with $h$ the determinant of the metric associated to (\ref{gTC}).

Finally, after considering all these points, we obtain the spectral density given by
\be  
\chi_{B=0}= \frac{ T^2\, \wn 2^\wn}{G_5
\left|{}_2F_1\left(1+\frac{1-i}{2}\wn,\, \frac{1-i}{2}\wn,\, 1-i\wn;\, \frac{1}{2}\right)\right|^2}
\,, \label{chib0}
\ee
which is identical to the expression (4.21) in \cite{Mateos:2007yp}  except for here the matter fields are in the adjoin representation and hence the result is proportional to $\nc^2$ trough $G_5=\frac{\pi}{2 \nc^2}$, while in (4.21) of \cite{Mateos:2007yp} it has to be proportional to $\nc\nf$, as seen in the constant $\tilde{\cal N}_\mt{D7}=\frac{1}{4}\nc\nf T^2$ there, which also accounts for the $T^2$ factor. To compare with (4.21) of \cite{Mateos:2007yp}, also notice that the change in the sign in the power at which the number 2 in the numerator of \eqref{chib0} is elevated compensates for the sign changes inside the argument of the hypergeometric function.

\begin{figure}[h!]
    \begin{center}
        \includegraphics[width=0.7\textwidth]{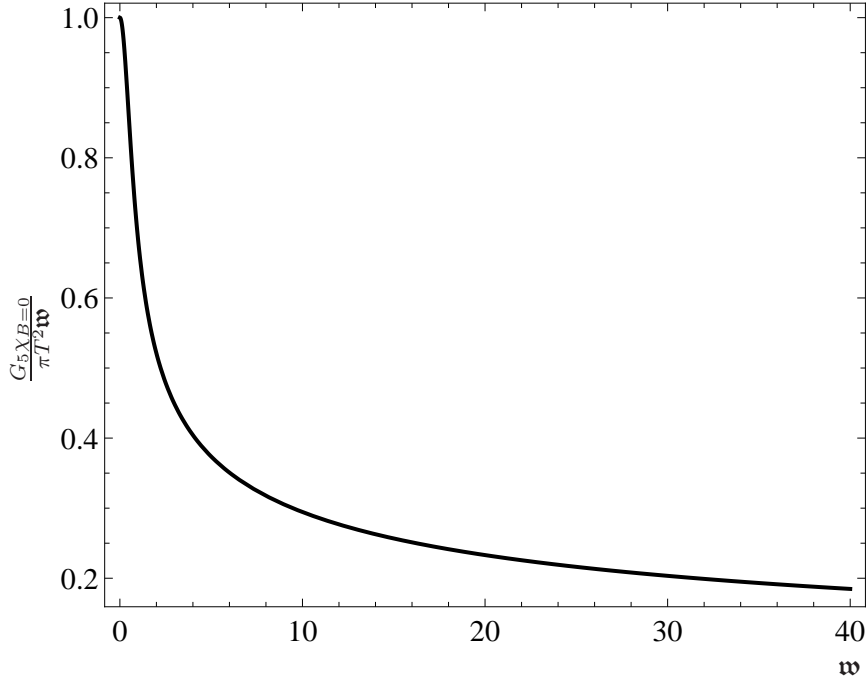}
         \put(-320,100){\rotatebox{90}{$\frac{G_5 \chi_{B=0}}{\pi T^2 \wn} $}}
         \put(-10,-10){$\wn$}
        \caption{The spectral density for $B=0$}
        \label{plotiso}
    \end{center}
\end{figure}

The dimensionless combination $\frac{G_5 \chi_{(1)}}{\pi T^2 \wn} $ is plotted in Fig. \eqref{plotiso}.

In the following sections we shall use these expressions as a reference for the results we will get with $B\neq 0$, so we turn now to the case with finite $B$, and for the sake of structure, we will study separately the cases of photons propagating parallel and perpendicular to the direction of the background field.

\subsection{Photons propagating in the direction of the background field}

When the momentum of the photon and the background field are aligned in the $z$ direction, we only need to compute the correlators $G^\mt{R}_{xx}$ and $G^\mt{R}_{yy}$, which, given the SO(2) symmetry in the $x$-$y$ plane, will be identical.

The procedure described here will be found in this section only, since the same steps have to be followed in the sections below, differing in the number of fields involved and some peculiarities of the solution, but the way to extrapolate should be clear.

We start by solving the field equations in the gravity side. As we will see below, for this propagation direction we only need the solution involving $A_x$ and $A_y$, and given the symmetry we have for this case this two are equal to each other. For definiteness we will work with $A_y$. The set of equations (\ref{eqAt}-\ref{eqyr}) that contains $A_y$ decouples further and reduces to 
\begin{multline}    \label{eomayZ1}
 U^2(r) V(r) W(r)\ A''_y(r) + U(r) V(r) \bigg[W(r) U'(r) + \frac{1}{2} U(r) W'(r) \bigg] A'_y \\
  + V(r) \Big[ k_t^2\ W(r) - k_z^2\ U(r) \Big]\ A_y + i\ B \Big[k_t\ W(r)\ h_{tx} - k_z\ U(r)\ h_{xz} \Big] =0
  \end{multline}
\begin{multline}   \label{eomayZ2}
 4\ B\  U(r) W(r)\ A'_y(r) +  i\ V(r) \Big[ k_t\ W(r)\ h'_{tx}(r) - k_z\ U(r)\ h'_{xz}(r) \Big] \\
 + i\ V'(r) \Big[ k_z\ U(r)\ h_{xz}(r) - k_t\ W(r)\ h_{tx}(r) \Big] = 0,
 \end{multline}
 \begin{multline} \label{eomayZ3}
U(r)V^2(r)W(r)\ h''_{tx}(r) + \frac{1}{2}\ U(r)V^2(r)W'(r)\ h'_{tx}(r) \\
- \bigg[ k_z^2\ V^2(r) + \frac{4}{3}\ B^2\ W(r) + 8\ V^2(r)W(r) - V(r)W(r)U'(r)V'(r) \Big]\ h_{tx}(r) \\
+ k_t\ k_z\ V^2(r)\ h_{xz}(r) + 4\ i\ B\ k_t\ V(r)W(r)\ A_y(r) = 0, 
\end{multline}
\begin{multline}   \label{eomayZ4}
U^2(r)V^2(r)W(r)\ h''_{xz}(r) + U(r)V^2(r) \bigg[ W(r)U'(r) - \frac{1}{2}\ U(r)W'(r) \bigg]\ h'_{xz}(r) \\
- \bigg[ \frac{4}{3}\ B^2\ U(r)W(r) - k_t^2\ V^2(r)W(r) + 8\ U(r)V^2(r)W(r) - U^2(r)V(r)V'(r)W'(r) \bigg]\ h_{xz}(r) \\
- k_t\ k_z\ V^2(r)W(r)\ h_{tx}(r) + 4\ i\ B\ k_z\ U(r)V(r)W(r)\ A_y(r) = 0.
\end{multline} 
 
These are four equations for the three fields $A_y, h_{tx}$ and $h_{zx}$, nonetheless, any solution to the first three equations will be a solution to the fourth, and we have checked this explicitly. Furthermore, we have solved separately for the two subset of equations, that is, the first three \eqref{eomayZ1}-\eqref{eomayZ3}, and  \eqref{eomayZ1}, \eqref{eomayZ2} and \eqref{eomayZ4}, obtaining the same solutions. 
Before proceeding any further, lets examine the space of solutions to \eqref{eomayZ1}-\eqref{eomayZ4}, which is the same as the space of solutions to the first three equations. 

First of all, if we study the large $r$ behavior, we find that the asymptotic solution for the $h(r)$ fields is $h(r)\sim r^2 \tilde{h}(r)$, where $\tilde{h}(r)$ approaches a constant as $r$ goes to infinity. To follow the procedure developed in \cite{Kaminski:2009dh} that we describe bellow for our particular case, we need to work with the $\tilde{h}(r)$ fields, so in practice we will use \eqref{eomayZ1}-\eqref{eomayZ4} as the equations for $A$ and $\tilde{h}(r)$. To keep the notation clean, from now on we will refer to the fields normalized in this way as $h$ and hope for no confusion to arise.

Going back to equations \eqref{eomayZ1}-\eqref{eomayZ3}, we notice that two of them are second order differential equations while the remaining one is first order, so the space of solutions is five dimensional. The equations are singular at the horizon, and using the Frobenius method to write the indicial polynomial we discover that close to the horizon there are two infalling solutions where all the fields behave as $(r-r_h)^{-\alpha}$, one regular for which the fields approach a constant, and two outgoing that behave as $(r-r_h)^\alpha$, with $\alpha$ given by the usual result $\alpha= i \frac{\wn}{2}$, $\wn=\frac{k_0}{2\pi T}$, where $T$ is the temperature associated to all the members of our family of backgrounds and $k_0$ is the time component of the photon momentum vector. Since our intention is to determine the retarded Green function $G^\mt{R}_{yy}$, we need to discard the two outgoing solutions, evaluate the boundary action on the remaining space of solutions and vary it with respect to the value that $A_y$ adopts at the boundary.

To see this program through, and given that the equations of motion are coupled, we need to follow \cite{Kaminski:2009dh} and find the three unique combinations of the two in falling solutions with the regular one that provide the three solutions that close to the boundary behave as
\be
\lim_{r \to \infty}
\begin{pmatrix}
A^1_y(r) \\
h^1_{tx}(r) \\
h^1_{zx}(r)
\end{pmatrix}
=
\begin{pmatrix}
1 \\
0 \\
0
\end{pmatrix},\,\, \lim_{r \to \infty}
\begin{pmatrix}
A^2_y(r) \\
h^2_{tx}(r) \\
h^2_{zx}(r)
\end{pmatrix}
=
\begin{pmatrix}
0 \\
1 \\
0
\end{pmatrix} \quad \text{and} \quad
\lim_{r \to \infty}
\begin{pmatrix}
A^3_y(r) \\
h^3_{tx}(r) \\
h^3_{zx}(r)
\end{pmatrix}
=
\begin{pmatrix}
0 \\
0 \\
1
\end{pmatrix}, \label{largeRbeh}
\ee
where the superscripts in the fields label the three independent solutions to the field equations. Given that the equations of motion are linear, their general, not outgoing, solution can be written as
\be
\mathbf{Sol}=\begin{pmatrix}
A^S_y(r) \\
h^S_{tx}(r) \\
h^S_{zx}(r)
\end{pmatrix}
=A_y^{bdry} \begin{pmatrix}
A^1_y(r) \\
h^1_{tx}(r) \\
h^1_{zx}(r)
\end{pmatrix}
+h_{tx}^{bdry} \begin{pmatrix}
A^2_y(r) \\
h^2_{tx}(r) \\
h^2_{zx}(r)
\end{pmatrix}
+h_{zx}^{bdry}\begin{pmatrix}
A^3_y(r) \\
h^3_{tx}(r) \\
h^3_{zx}(r)
\end{pmatrix}, \label{gensol}
\ee
where $A_y^{bdry}$ is the value of the $A_y$ field at the boundary, making this way of writing the general solution particularly suitable to take variations with respect to that quantity.

Now we can evaluate the boundary action on this general solution and vary it with respect to $A_y^{bdry}$ twice, since we are looking for the two point correlation function. The values of the fields at the boundary are independent, so the variation with respect to $A_y^{bdry}$ of either $h^S\vert_{bdry}$ field evaluated at the boundary is zero, but as far as the derivatives are concerned, we see from differentiating (\ref{gensol}) with respect to $r$ that the variations are
\be
\dfrac{\delta{A_y^S}'\vert_{bdry}}{\delta {A_y}^{bdry}} = {{A_y^{1}}'} \vert_{bdry} \,\,\,\,\ , \,\,\,\,\ \dfrac{\delta{{h_{tx}^S}'}\vert_{bdry}}{\delta {A_y}^{bdry}} = {{h_{tx}^{1}}'} \vert_{bdry} \,\,\,\,\ \text{and}\,\,\,\,\ \dfrac{\delta{{h_{zx}^S}'}\vert_{bdry}}{\delta {A_y}^{bdry}} = {{h_{zx}^{1}}'} \vert_{bdry} . \label{varSolAy}
\ee
We can use \eqref{varSolAy} to vary the on shell action with respect to $A_y^{bdry}$, and doing so twice to obtain ${G^\mt{R}}^{yy}$ we notice that the only term that does not vanish is 
\be
{G^\mt{R}}^{yy}(k)= -\frac{2}{8\, \pi\, G_5} \left[ U(r)\sqrt{W(r)} {{A_y^{1}}'}\right]_{bdry} . \label{greenyyZ}
\ee
As we see, the problem boils down to finding ${A_y^{1}}'$ and evaluating the product in (\ref{greenyyZ}) at the boundary.

To find the solutions with the behavior described in (\ref{largeRbeh}), we need to find three arbitrary linear independent solution of the equations of motion and use them as the columns of a matrix,
\be
{\cal M}=\begin{pmatrix}
{A_y^{a}}(r) & {A_y^{b}}(r) & {A_y^{c}}(r) \\
{h_{tx}^{a}}(r) & {h_{tx}^{b}}(r) & {h_{tx}^{c}}(r) \\
{h_{zx}^{a}}(r) & {h_{zx}^{b}}(r) & {h_{zx}^{c}}(r)
\end{pmatrix}.
\ee
Once the matrix ${\cal M}$ has been defined we see that 
\be
\begin{pmatrix}
{A_y^{1}}(r) & {A_y^{2}}(r) & {A_y^{3}}(r) \\
{h_{tx}^{1}}(r) & {h_{tx}^{2}}(r) & {h_{tx}^{3}}(r) \\
{h_{zx}^{1}}(r) & {h_{zx}^{2}}(r) & {h_{zx}^{3}}(r)
\end{pmatrix}={\cal M} ({\cal M}^{-1}\vert_{bdry}).
\ee

As mentioned before, the metric components appearing in the equations of motion are only known numerically for generic values of $B/V_0$, so we had to recur to numerical methods to find the solution to the field equations. In practice, to construct the solutions $a, b$ and $c$, we used Frobenius method to find the perturbative solutions around the horizon, chose the two infalling and the regular one, and integrated out towards the boundary.

\begin{figure}[h!]
    \begin{center}
        \includegraphics[width=0.7\textwidth]{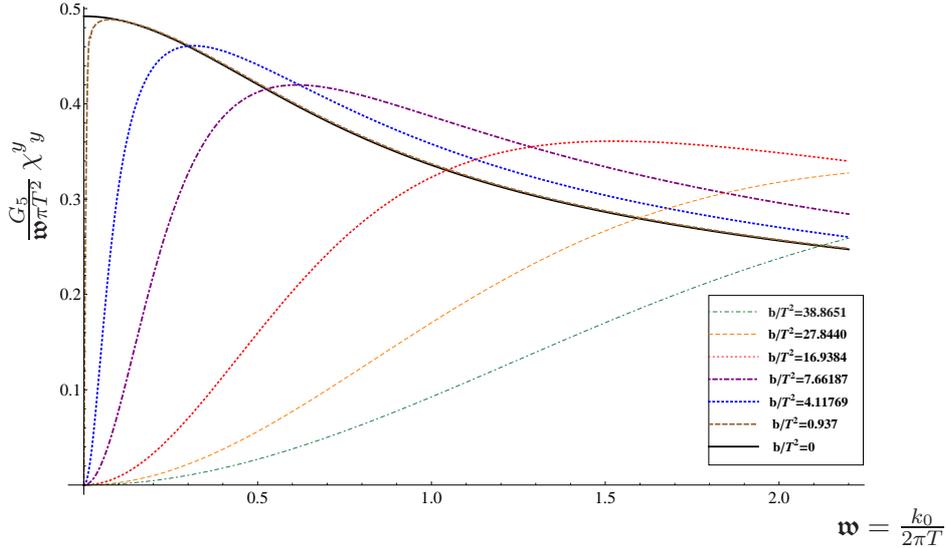}
         \put(-320,100){\rotatebox{90}{$\frac{G_5}{\wn \pi T^2}\, \chi^y_{\ y}$}}
         \put(-10,-10){$\wn=\frac{k_0}{2\pi T}$}
        \caption{Spectral density ${\chi^y}_{y}$ when the photon momentum is aligned with the background field $B$. Note that when $\wn = 0$ the DC conductivity drops down to zero for non-vanishing values of the magnetic field.}
        \label{chiyyZ}
    \end{center}
\end{figure}

Below we present the results of these calculations, starting in Fig. \eqref{chiyyZ} with the spectral density ${\chi^y}_{y}(\wn)=-2 \mbox{ Im } {{G^{\mt{R}}}^y}_{y}(\wn)$, which will be the same as ${\chi^x}_{x}$, for a number of intensities of $B$. Notice that since $F = B\ dx\wedge dy$, the intensity of the physical field $b$ will be scaled as $b=B/C_1$, and the results will be reported for the values of the dimensionless quantity $b/T^2$. Unless specified otherwise, the color and form line code for the different values of $b/T^2$ in this and all the subsequent figures is as follows: black $b/T^2=0$, brown (thick, dashed) $b/T^2=0.937$, blue (thick, dotted) $b/T^2=4.118$, purple (thick, dot-dashed) $b/T^2=7.662$, red (thin, dotted) $b/T^2=16.938$, orange (thin,dashed) $b/T^2=27.844$, green (thin, dot-dashed) $b/T^2=38.865$. 

\begin{figure}[h!]
    \begin{center}
        \includegraphics[width=0.7\textwidth]{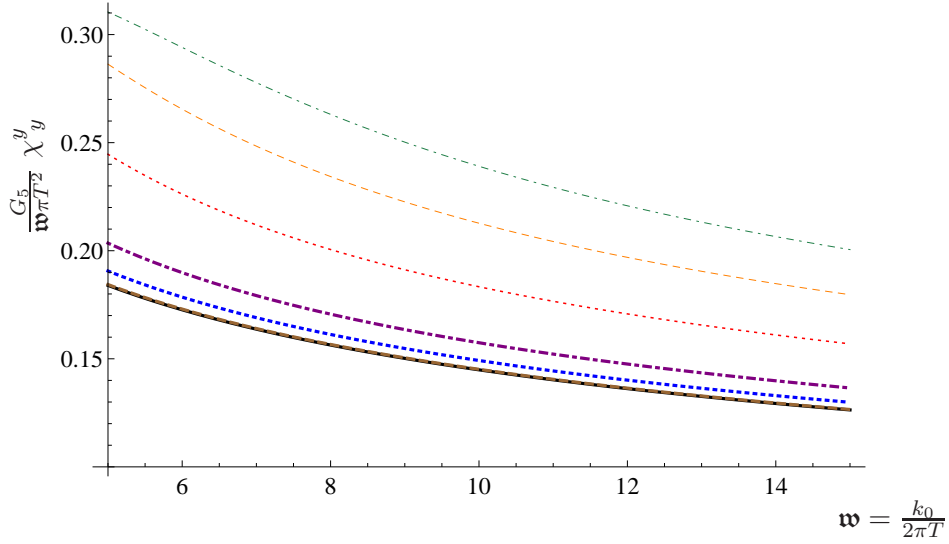}
         \put(-320,100){\rotatebox{90}{$\frac{G_5}{\wn \pi T^2}\ \chi^y_{\ y}$}}
         \put(-10,-10){$\wn=\frac{k_0}{2\pi T}$}
        \caption{Spectral density ${\chi^y}_{y}$ when the photon momentum is aligned with the background field $B$ for large frequencies.}
        \label{chiyyZlargeW}
    \end{center}
\end{figure}

In Fig. \eqref{chiyyZlargeW} we show the same quantity for large values of $\wn$.

\begin{figure}[h!]
    \begin{center}
        \includegraphics[width=0.7\textwidth]{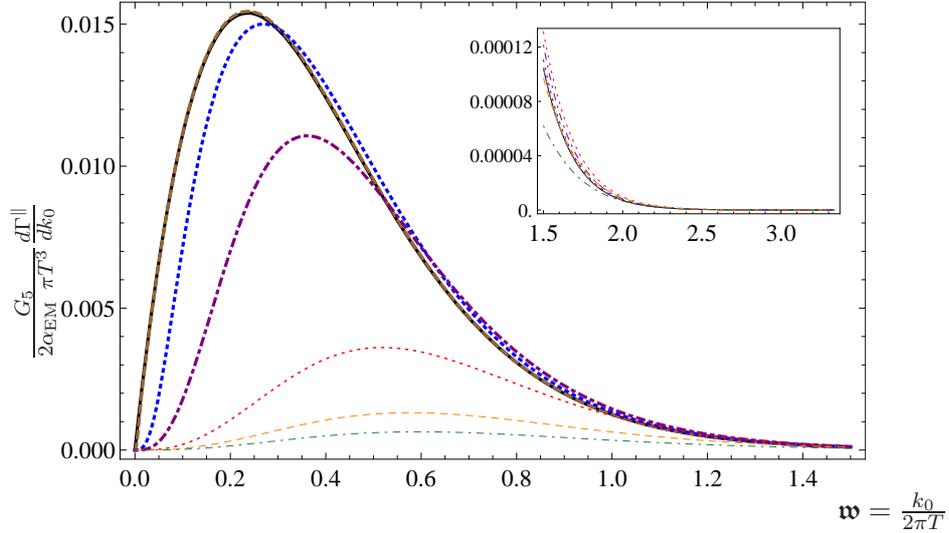}
         \put(-320,50){\rotatebox{90}{$\frac{G_5}{2 \alpha_{\textrm{EM}}\ \pi T^3} \frac{d \Gamma^{\scriptscriptstyle{\parallel}}}{d k_0}$}}
         \put(-10,-10){$\wn=\frac{k_0}{2\pi T}$}
        \caption{Total production rate when the photon momentum is aligned with the background field $B$. Detail for large frequencies in the insert.}
        \label{TotZ}
    \end{center}
\end{figure}

Now that we have obtained ${\chi^y}_{y}(\wn)$ and in doing so, also ${\chi^x}_{x}(\wn)$, we can plot the total production rate given by (\ref{diff}) with the appropriate polarization vectors corresponding to $\vartheta = 0$ and $s=\{1,2\}$. We convert this quantity to emission rate per unit photon energy in an infinitesimal angle around $\vartheta=0$. Using that the photon momentum is light-like we get 
\be
\frac{ G_5}{2\alpha_\mt{EM} T^3}\frac{d\Gamma}{d(\cos\vartheta)\, d k^0}=\frac{ G_5 \wn}{2 T^2}\frac{1}{e^{2\pi \wn}-1}\left(\chi_{(1)}+\chi_{(2)}\right)\,
\ee
This quantity for different values of $b/T^2$ is plotted in Fig. \eqref{TotZ}. 

We can use (\ref{chib0}) to get the result for $B=0$ at the same temperature 
\be
\frac{ G_5}{2\alpha_\mt{EM} T^3}\frac{d\Gamma_\mt{iso}}{d(\cos\vartheta)\, d k^0}=\frac{\wn^2 2^\wn}{\left(e^{2\pi \wn}-1\right)
\left|{}_2F_1\left(1+\frac{1-i}{2}\wn_s,\, \frac{1-i}{2}\wn_s,\, 1-i\wn_s;\, \frac{1}{2}\right)\right|^2}\,,
\ee 
which is identical to the result in \cite{Mateos:2007yp} and is plotted as a black dotted curve in the figure along with the plots for different values of $b/T^2$.  

Note that at large $k_0$ (more specifically, for $k_0\gg b$) the spectral density does not converge to the same constant as the $B=0$ case does. We can explain this recalling that our results are dependent on two independent dimensionless ratios: Taking $k_0\to \infty$ while the value of $b$ and $T$ are kept fixed is not equivalent to taking $b\to 0$ with $k_0$ and $T$ fixed, this is so because $b/T^2$ remains finite when the first limit is intended.

The value of the spectral function for $\wn=0$ in Fig. \eqref{chiyyZ} provides the DC conductivity in the direction perpendicular to $B$, and we see that it drops down to zero for any non vanishing value of $B$. In the next section we will see that this is not the case for the conductivity along the direction in which $B$ is pointed, and we will depict its behavior with respect to the intensity of $B$. All these results will be discussed in the analysis section.


\subsection{Photons propagating perpendicularly to the background field}

To study the production of photons propagating perpendicularly to the background field we will use the expressions of section (\ref{sec1}) with $\vartheta=\pi/2$, i.e., the momentum of the photon points in the $x$ direction.

The steps of the procedure are the same as in the previous section, so we will just state the particulars.

In this case we need to compute separately $G^\mt{R}_{yy}$ and $G^\mt{R}_{zz}$ as they and the equations that define them are different since there is no rotational symmetry in the $y$-$z$ plane.

Let us start with the set of equations (\ref{eqAy}-\ref{eqrr}), which in this case decouples $A_z$ further into
\begin{multline}
U^2(r) V^2(r) W(r)\ A''_z(r) \\
+ U(r) V(r) \bigg[ V(r) W(r) U'(r) + U(r) W(r) V'(r) - \frac{1}{2}\ U(r) V(r) W'(r) \bigg]\ A'_z(r)\\
 + V(r) W(r)\ \Big[ k_t^2\ V(r) - k_x^2\ U(r) \Big]\ A_z(r) - i\ B\ k_x\ U(r) W(r)\ h_{yz}(r) = 0,
\end{multline}
\begin{multline}
U^2(r) V^2(r) W(r)\ h''_{yz}(r) + U(r) V^2(r) \bigg[ W(r) U'(r) - \frac{1}{2} U(r) W'(r) \bigg]\ h'_{yz}(r) \\
+ \bigg[ k_t^2\ V^2(r) W(r) -  k_x^2\ U(r) V(r) W(r) - 8\ U(r) V^2(r) W(r) \\ 
+ U^2(r) V(r) V'(r) W'(r) - \frac{4}{3}\ B^2\ U(r) W(r) \bigg]\ h_{yz}(r) \\
+ 4\ i\ B\ k_z\ U(r) V(r) W(r)\ A_z(r) = 0, 
\end{multline}
which are two second order differential equations involving only the fields $A_z$ and $h_{yz}$.

We again will need to extract an $r^2$ factor from $h_{yz}$, and in this case the family of solutions will be four dimensional, consisting of two infalling and two outgoing solutions, showing the same near horizon behavior $(r-r_h)^{\pm\alpha}$ with $\alpha$ as in the previous section. 

To construct expressions similar to (\ref{largeRbeh}) and (\ref{gensol}) we need only two solutions, and we will use the infalling ones. The variation of the on shell action twice with respect to ${A_z}^{bdry}$ leads us to one non vanishing term providing ${G^\mt{R}}^{zz}$ as 
\be
{G^\mt{R}}^{zz}(k)=  -\frac{2}{8\ \pi\ G_5} \left[ \frac{U(r)V(r)}{\sqrt{W(r)}} {{A_z^{1}}'}\right]_{bdry} , \label{greenzzX}
\ee
where ${A_z^{1}}$ is the part of the solution equivalent to ${A_y^{1}}$ found in (\ref{largeRbeh}) and (\ref{gensol}).

The results of this calculation are shown in the figures below, starting with Fig. \eqref{chizzX} displaying the spectral density ${\chi^z}_{z}(\wn)=-2 \mbox{ Im } {{G^\mt{R}}^z}_{z}(\wn)$ for a series of values of $b/T^2$.

\begin{figure}[h!]
    \begin{center}
        \includegraphics[width=0.7\textwidth]{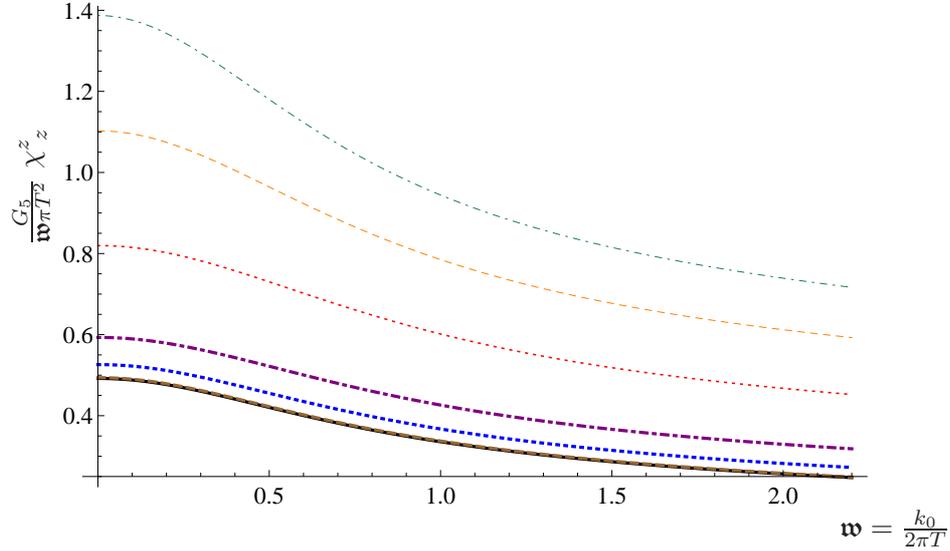}
         \put(-320,100){\rotatebox{90}{$\frac{G_5}{\wn \pi T^2}\ \chi^z_{\ z}$}}
         \put(-10,-10){$\wn=\frac{k_0}{2\pi T}$}
        \caption{Spectral density ${\chi^z}_{z}$ when the photon momentum is aligned perpendicular to the background field $B$. Note that when $\wn = 0$ the DC conductivity shows a non-trivial dependency on the intensity of the magnetic field.}
        \label{chizzX}
    \end{center}
\end{figure} 

\begin{figure}[h!]
    \begin{center}
        \includegraphics[width=0.7\textwidth]{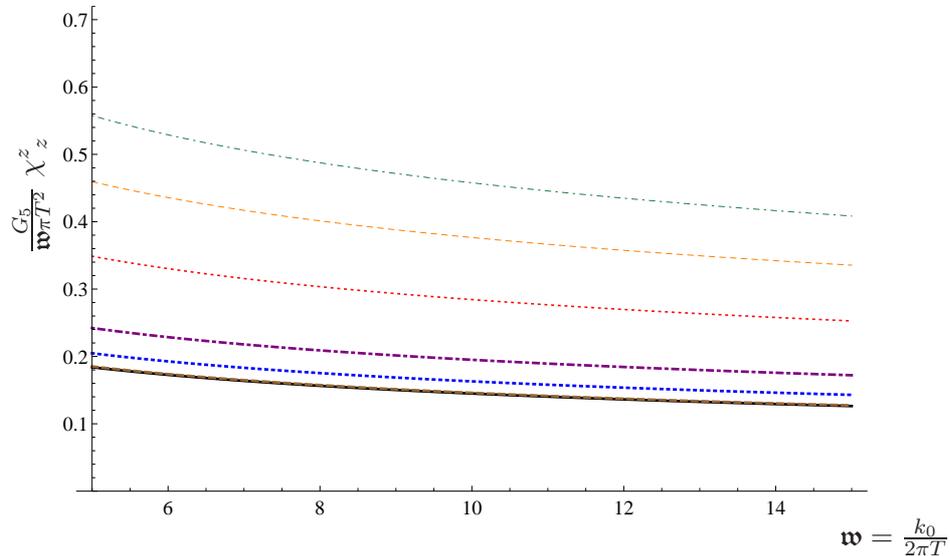}
         \put(-320,100){\rotatebox{90}{$\frac{G_5}{\wn \pi T^2}\ \chi^z_{\ z}$}}
         \put(-10,-10){$\wn=\frac{k_0}{2\pi T}$}
        \caption{Spectral density ${\chi^z}_{z}$ when the photon momentum is aligned perpendicular to the background field $B$, for large frequencies.}
        \label{chizzXlargeW}
    \end{center}
\end{figure}

In Fig. \eqref{chizzXlargeW} we show the same quantity for large values of $\wn$.

\begin{figure}[h!]
    \begin{center}
        \includegraphics[width=0.7\textwidth]{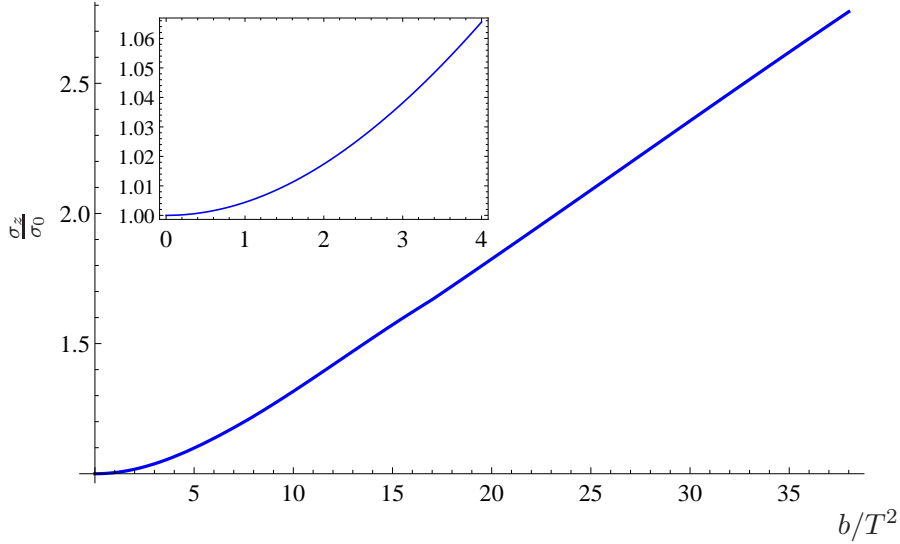}
         \put(-320,100){\rotatebox{90}{$\frac{\sigma_z}{\sigma_0}$}}
         \put(-10,-10){$b/T^2$}
        \caption{Normalized DC conductivity in the direction of the magnetic field as a function of $b/T^2$. Details for small values of $b/T^2$ in the insert.}
        \label{condZ}
    \end{center}
\end{figure}

In this case, the value of the spectral function for $\wn=0$ in Fig. \eqref{chizzX} provides the DC conductivity $\sigma_z$ in the direction of the field $B$. We see that it has a non trivial behavior with respect to the intensity of the magnetic field, so we depict this in Fig. \eqref{condZ}, normalized with respect to the value of the conductivity $\sigma_0$ for the case $B=0$.

Now, we turn to the set of equations (\ref{eqAt}-\ref{eqrr}), which for this propagation direction, decouples $A_y$ further into
\begin{multline} \label{eqkxAy}
U^2(r) V^2(r) W(r)\ A''_y(r) + U(r) V^2(r) \Big[ W(r) U'(r) + \frac{1}{2} U(r) W'(r) \Big]\ A'_y(r) \\
- V(r) W(r) \Big[k_x^2\ U(r) - k_t^2\ V(r) \Big]\ A_y(r) + \frac{1}{2}\ i\ B\ k_x U(r) \Big[ V(r) h_{zz}(r) \\
- W(r)\ \Big(h_{xx}(r) + h_{yy}(r) \Big) \Big] + V(r) W(r) \Big[ k_x\ h_{tt}(r) + 2\ k_t\ h_{tx}(r) \Big] = 0,
\end{multline}
\begin{multline}  \label{eqkxXX}
U^2(r)V^2(r)W^ 2(r)\ h''_{xx}(r)  \\ 
+ \frac{1}{2}\ U(r)V(r)W(r) \bigg[ 2\ V(r)W(r)U'(r) - U(r)W(r)V'(r) + U(r)V(r)W'(r) \bigg]\ h'_{xx}(r) \\
+ \frac{1}{2}\ U(r)V(r)W(r)V'(r) \bigg[ V(r) W(r)\ h'_{tt}(r) + U(r)W(r)\ h'_{yy}(r) + U(r)V(r)\ h'_{zz}(r) \bigg] \\
- V^2(r)W^2(r) \bigg[ k_x^2 + \frac{1}{2}\ U'(r)V'(r) \bigg]\ h_{tt}(r) + W^2(r) \bigg[ k_t^2\ V^2(r) - 8\ U(r)V^2(r) + \frac{1}{2}\ U^2(r)V'^2(r) \bigg] h_{xx}(r) \\
- U(r)W^2(r) \bigg[ \frac{8}{3}\ B^2 + k_x^2\ V(r) + \frac{1}{2}\ U(r)V'^2(r) \bigg] h_{yy}(r) - U(r)V^2(r) \bigg[k_x^2\ W(r) + \frac{1}{2}\ U(r)V'(r) W'(r) \bigg] h_{zz}(r) \\
- 2\ k_t\ k_x\ V^2(r)W^2(r)\ h_{tx}(r) + \frac{16}{3}\ i\ B\ k_x\ U(r)V(r)W^2(r)\ A_y(r) = 0,
\end{multline}
\begin{multline} \label{eqkxYY}
U^2(r)V^2(r)W^2(r)\ h''_{yy}(r)  \\ 
+ \frac{1}{2}\ U(r)V(r)W(r) \bigg[2\ V(r)W(r)U'(r) - U(r)W(r)V'(r) + U(r)V(r)W'(r) \bigg]\ h'_{yy}(r) \\
+ \frac{1}{2}\ U(r)V(r)W(r)V'(r) \bigg[V(r)W(r)\ h'_{tt}(r) + U(r)W(r)\ h'_{xx}(r) + U(r)V(r)\ h'_{zz}(r) \bigg] \\
- \frac{1}{2}\ V^2(r)W^2(r)U'(r)V'(r)\ h_{tt}(r) - \frac{1}{6}\ U(r)W^2(r) \bigg[ 16\ B^2 + 3\ U(r)V'^2(r) \bigg]\ h_{xx}(r) \\
-  W^2(r) \bigg[ k_x^2\ U(r)V(r) - k_t^2\ V^2(r) + 8\ U(r)V^2(r) - U^2(r)V'^2(r) \bigg]\ h_{yy}(r) \\
- \frac{1}{2}\ U^2(r)V^2(r)V'(r)W'(r)\ h_{zz}(r) + \frac{16}{3}\ i\ B\ k_x\ U(r)V(r)W^2(r)\ A_y(r) = 0,
\end{multline} 
\begin{multline}  \label{eqkxZZ}
U^2(r)V^3(r)W^2(r)\ h''_{zz}(r) \\
+ U(r)V^2(r)W(r) \bigg[ V(r)W(r)U'(r) + U(r)W(r)V'(r) - U(r)V(r)W'(r) \bigg]\ h'_{zz}(r) \\
+ \frac{1}{2}\ U(r)V^2(r)W(r)W'(r) \bigg[ V(r)W(r)\ h'_{tt}(r) + U(r)W(r)\ h'_{xx}(r) - 2\ U(r)V(r)\ h'_{zz}(r) \bigg] \\
- \frac{1}{2}\ V^3(r)W^2(r)U'(r)W'(r)\ h_{tt}(r)  + U(r)W^2(r) \bigg[ \frac{4}{3}\ B^2\ W(r) - \frac{1}{2}\ U(r)V(r)V'(r)W'(r) \bigg]\ \Big( h_{xx}(r) + h_{yy}(r) \Big) \\
- V(r) \bigg[ \frac{4}{3}\ B^2 U(r)W^2(r) + k_x^2\ U(r)V(r)W^2(r) - k_t^2 V^2(r)W^2(r) \\ 
+ 8\ U(r)V^2(r)W^2(r) - \frac{1}{2}\ U^2(r)V^2(r)W'^2(r) \bigg]\ h_{zz}(r) - \frac{8}{3}\ i\ B\ k_x\ U(r)V(r)W^3(r)\ A_y(r) = 0,
\end{multline}
\begin{multline}  \label{eqkxRR}
U^2(r)V^2(r)W^2(r) \bigg[V(r)W(r)\ h''_{tt}(r) + U(r)W(r)\ h''_{xx}(r) +  U(r)W(r)\ h''_{yy}(r) + U(r)V(r)\ h''_{zz}(r) \bigg] \\
- \frac{1}{2}\ U(r)V^3(r)W^3(r)U'(r)\ h'_{tt}(r) + U^2(r)V(r)W^3(r) \bigg[ \frac{1}{2}\ V(r)U'(r) -  U(r)V'(r) \bigg] \Big( h'_{xx}(r) + h'_{yy}(r) \Big) \\
+ U^2(r)V^3(r)W(r) \bigg[ \frac{1}{2}\ W(r)U'(r) + U(r)W'(r) \bigg]\ h'_{zz}(r) + V^3(r)W^3(r) \bigg[\frac{1}{2}\ U'^2(r) - U(r)U''(r) \bigg] h_{tt}(r) \\
+ U^2(r)W^3(r) \bigg[ \frac{4}{3}\ B^2 - \frac{1}{2}\ V(r)U'(r)V'(r) + U(r)V'^2(r) - U(r)V(r)V''(r) \bigg] \Big( h_{xx}(r) + h_{yy}(r) \Big)  \\
+ U^2(r)V^3(r) \bigg[ U(r)W'^2(r) - U(r)W(r)W''(r) - \frac{1}{2}\ W(r)U'(r)W'(r) \bigg]\ h_{zz}(r) \\
- \frac{8}{3}\ i\ B\ k_x\ U^2(r)V(r)W^3(r)\ A_y(r) = 0,
\end{multline}
\begin{multline} \label{eqkxTR}
U(r)V(r)W^2(r) \bigg[k_x\ h'_{tx} - k_t \big( h'_{xx}(r) + h'_{yy}(r) \big) \bigg] - k_t\ U(r)V^2(r)W(r)\ h'_{zz}(r) \\
- k_x\ V(r)W^2(r)U'(r)\ h_{tx}(r) + k_t\ W^2(r) \bigg[ V(r)U'(r) + \frac{1}{2} U(r)V'(r) \bigg] h_{xx}(r) \\
+ \frac{1}{2}\ k_t\  W^2(r)\ \Big[U(r)V(r)\Big]'\  h_{yy}(r)+  \frac{1}{2}\ k_t\ V^2(r)\ \Big[U(r)W(r)\Big]'\ h_{zz}(r) = 0,
\end{multline}
\begin{multline}  \label{eqkxTT}
U^2(r) V^3(r) W^2(r)\ h''_{tt}(r) + \frac{1}{2}\ U^2(r)V^2(r)W(r) U'(r) \Big[ W(r)\Big( h'_{xx}(r) + h'_{yy}(r) \Big) + V(r)\ h'_{zz}(r) \Big] \\
+ U(r)V^2(r) W(r) \bigg[U(r) W(r) V'(r) + \frac{1}{2}\bigg( U(r) V(r) W'(r) - V(r)W(r)U'(r)\bigg)  \bigg]\ h'_{tt}(r) \\
+ V(r)W^2(r) \bigg[ \frac{1}{2}\ V^2(r)U'^2(r) - \frac{4}{3}\ B^2\ U^2(r) - 6\ k_x^2\ U(r)V(r) - 8\ U(r)V^2(r) \bigg]\  h_{tt}(r)  \\
+ U(r)W^2(r) \bigg[ \frac{4}{3}\ B^2\ U(r) + k_t^2\ V^2(r) - \frac{1}{2}\ U(r)V(r)U'(r)V'(r) \bigg]\ \Big(h_{xx}(r) + h_{yy}(r) \Big) \\
+ U(r)V^3(r) \bigg[ k_t^2\ W(r) - \frac{1}{2}\ U(r)U'(r)W'(r) \bigg] h_{zz}(r) \\ 
- 2\ k_t k_x\ U(r) V^2(r) W^2(r)\ h_{tx}(r) - \frac{8}{3}\ i\ B\ k_x\ U^2(r)V(r)W^2(r)\ A_y(r) = 0,
\end{multline}
\begin{multline} \label{eqkxTX}
U(r)V^2(r)W(r)\ h''_{tx}(r) + \frac{1}{2}\ U(r)V^2(r)W'(r)\ h'_{tx}(r) \\ 
 - k_t\ k_x\ V(r) \Big[ V(r)\ h_{zz}(r) + W(r)\ h_{yy}(r) \Big] + W(r) \bigg[ V(r)U'(r)V'(r) - \frac{4}{3}\ B^2 - 8\ V^2(r) \bigg]\  h_{tx}(r) \\
  + 4\ i\ B\ k_t\ V(r)W(r)\ A_y(r) = 0,  
\end{multline}
\begin{multline}  \label{eqkxXR}
U(r)V^2(r)W^2(r) \Big[ k_x\ h'_{tt}(r) + k_t\ h'_{tx}(r) \Big] + k_x\ U^2(r)V(r)W(r) \Big[ W(r)\ h'_{yy}(r) + V(r)\ h'_{zz}(r) \Big] \\
 - \frac{1}{2}\ k_x\ V(r)W^2(r) \Big[U(r)V(r)\Big]'\ h_{tt}(r) - \frac{1}{2}\ k_x\ U^2(r)V(r) \Big[V(r)W(r)\Big]'\ h_{zz}(r) \\
- U(r)W^2(r)V'(r) \Big[ k_x\ U(r)\ h_{yy}(r) + k_t\ V(r)\ h_{tx}(r)\Big] - 4\ i\ B\ U^2(r)V(r)W^2(r)\ A'_y(r) = 0.
\end{multline}
These are nine equations involving only the six fields $A_y, h_{tt}, h_{xx}, h_{yy}, h_{zz}$ and $h_{tx}$. Again, we have verified that this system is redundant and in the order in which we list them, it is true that the solutions to the first six equations verify the last three. As in the previous case we solved more than one subset of six equations obtaining the same results.

Order counting is trickier with this system, but after the dust settles, we are left with three infalling and three outgoing solutions with the same exponent $\alpha$ as in the previous cases, plus two regular solutions that approach a constant at the horizon and one further regular solution that goes like $(r-r_h)^{1/2}$. Of this nine solutions we shall discount the outgoing ones and proceed with the remaining six in a similar fashion as we have done for the previous two cases. Once we variate the on shell action twice with respect to ${A_y}^{bdry}$ we are left with one non vanishing term that dictates ${G^\mt{R}}^{yy}$ for photons propagating along the $x$ direction as 
\be
{G^\mt{R}}^{yy}(k)= -\frac{2}{8\ \pi\ G_5} \left[ U(r)\sqrt{W(r)} {{A_y^{1}}'}\right]_{bdry} , \label{greenyyX}
\ee
which is identical in appearance to (\ref{greenyyZ}), but the difference is highly relevant and encoded in the values that ${{A_y^{1}}'}$ takes at the boundary for photons propagating along the direction $x$.

These results are shown in the following figures, beginning with Fig. \eqref{chiyyX} that displays the spectral density ${\chi^y}_y(\wn)$ for various values of the background field $B$. In this case no further information is obtained from the value of the spectral density at $\wn=0$, since this determines the DC conductivity in the $y$ direction, which was already learned in the previous section and coincides in showing that it drops to zero for non vanishing $B$.

\begin{figure}[h!]
    \begin{center}
        \includegraphics[width=0.7\textwidth]{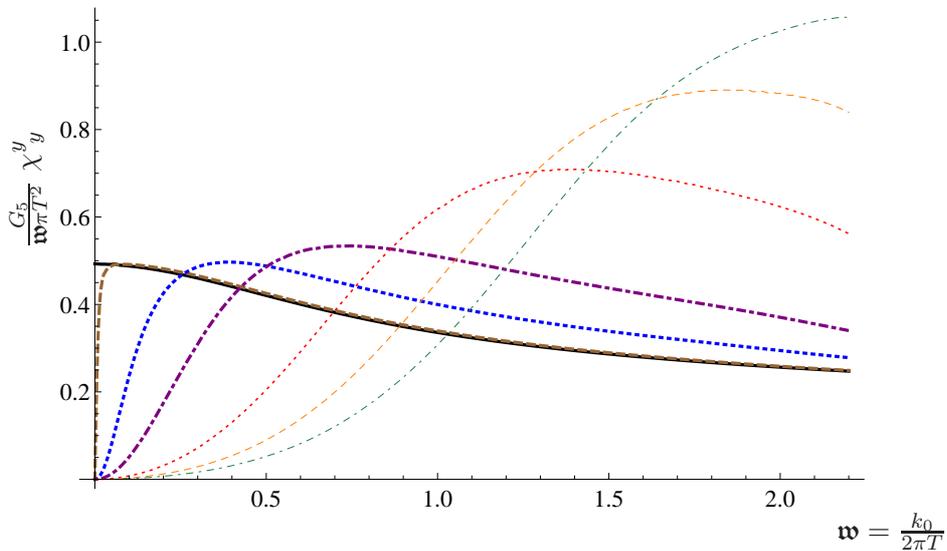}
         \put(-320,100){\rotatebox{90}{$\frac{G_5}{\wn \pi T^2}\ \chi^y_{\ y}$}}
         \put(-10,-10){$\wn=\frac{k_0}{2\pi T}$}
        \caption{Spectral density ${\chi^y}_{y}$ when the photon momentum is aligned perpendicular to the background field $B$. Note that when $\wn = 0$ the DC conductivity drops down to zero for non-vanishing values of the magnetic field.}
        \label{chiyyX}
    \end{center}
\end{figure} 

\begin{figure}[h!]
    \begin{center}
        \includegraphics[width=0.7\textwidth]{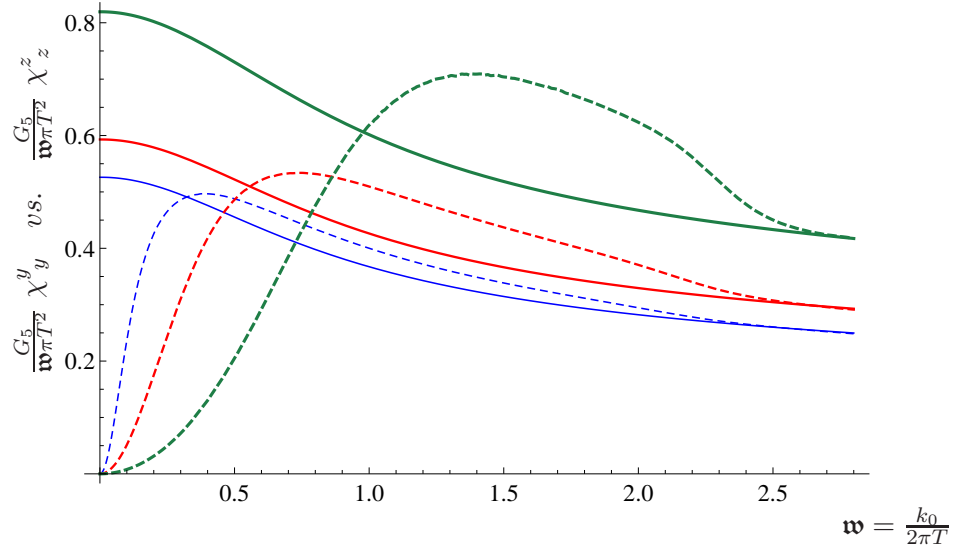}
         \put(-320,50){\rotatebox{90}{$\frac{G_5}{\wn \pi T^2}\ \chi^y_{\ y} \quad vs. \quad \frac{G_5}{\wn \pi T^2}\ \chi^z_{\ z}$}}
         \put(-10,-10){$\wn=\frac{k_0}{2\pi T}$}
        \caption{Comparison of the spectral densities ${\chi^y}_y$ (dashed lines) and ${\chi^z}_z$ (solid lines) for different values of $b/T^2$. The color code is: 1) blue $b/T^2= 4.118$, 2) red $b/T^2= 7.662$, 3) green $b/T^2= 16.939$. Notice that for each value of $b/T^2$ the two spectral densities converge for large values of $\wn$.}
        \label{comp}
    \end{center}
\end{figure}

In Fig. \eqref{comp} we compare the spectral densities ${\chi^y}_y$ and ${\chi^z}_z$ for a few values of $b/T^2$, and notice that for each of these, they converge for large values of $\wn$. Therefore Fig. \eqref{chizzXlargeW} also describes the large $\wn$ behavior for this polarization.

\begin{figure}[h!]
    \begin{center}
        \includegraphics[width=0.7\textwidth]{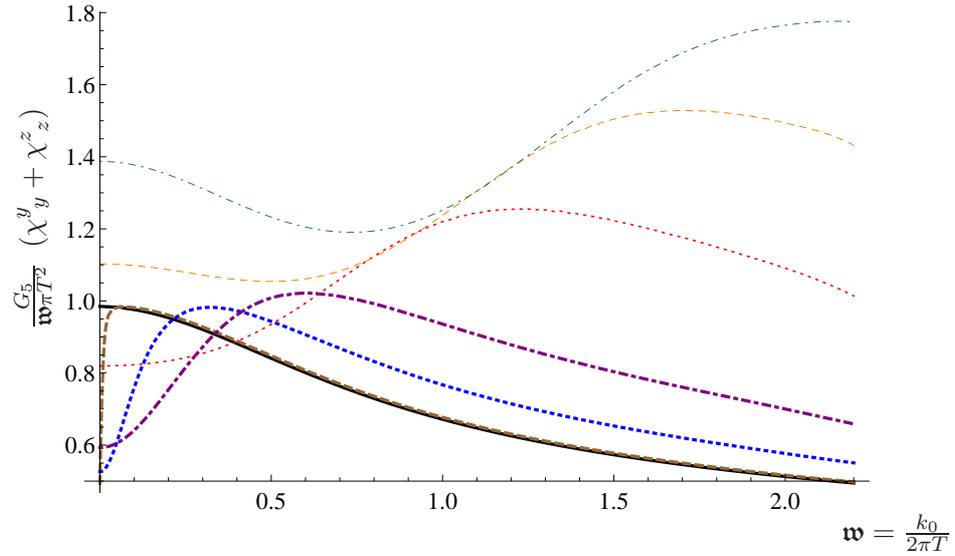}
         \put(-320,70){\rotatebox{90}{$\frac{G_5}{\wn \pi T^2}\ \left(\chi^y_{\ y}+\chi^z_{\ z}\right)$}}
         \put(-10,-10){$\wn=\frac{k_0}{2\pi T}$}
        \caption{Trace of the spectral function for photons propagating perpendicular to the background field $B$.}
        \label{TotX}
    \end{center}
\end{figure}

In Fig. \eqref{TotX} we report the trace of the spectral function for photons propagating perpendicular to the background field.

\begin{figure}[h!]
    \begin{center}
        \includegraphics[width=0.7\textwidth]{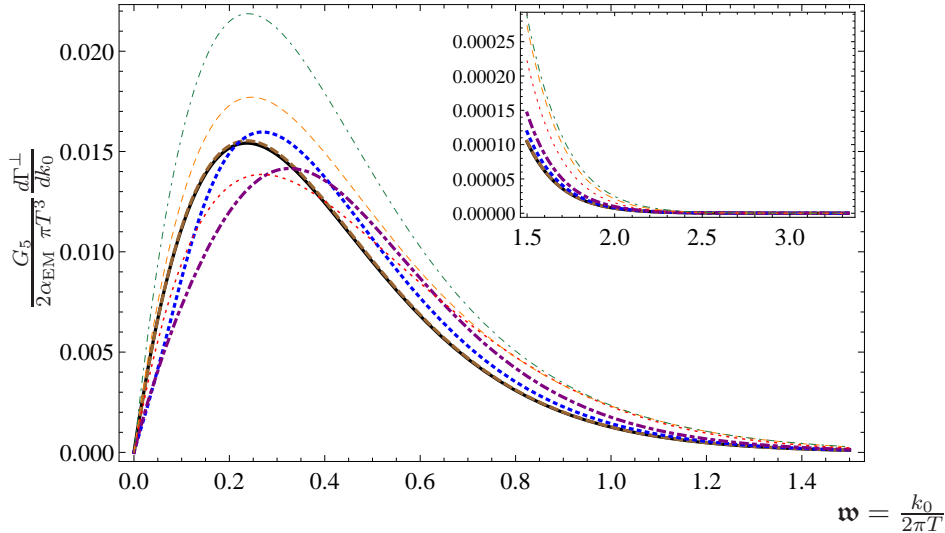}
         \put(-320,70){\rotatebox{90}{$\frac{G_5}{2 \alpha_{\textrm{EM}}\ \pi T^3} \frac{d \Gamma^{\scriptscriptstyle{\bot}}}{d k_0}$}}
         \put(-10,-10){$\wn=\frac{k_0}{2\pi T}$}
        \caption{Total production rate of photons propagating in the direction perpendicular to the background field $B$. In the insert we present the behavior for large frequencies $\wn$.}
        \label{gtot}
    \end{center}
\end{figure}

Now that we have the results for ${\chi^y}_y$ and ${\chi^z}_z$ for photons propagating in the $x$ direction, we are ready to add them and present the total production rate of this kind of photons, so we convert this quantity to emission rate per unit photon energy in an infinitesimal angle around $\vartheta=\pi/2$ and in Fig. \eqref{gtot} we report the plots for different values of $b/T^2$, including $B=0$ for comparison with previous work. 

\section{Analysis}   \label{analysis}

There are a few things that are worth stressing about the results that can be appreciated in the plots of the previous section.

As noticed in section \ref{sec2}, to consistently study the perturbations of the electromagnetic field in the presence of a finite magnetic background field, it is mandatory to take into account the contribution to the stress energy tensor coming from these perturbations. In a previous version of this article, just like in the references cited here, we had not considered this contribution, and that was the reason to replace the original manuscript.

Looking at the results of the calculations now that they have been done consistently, we see that the details of the spectral functions are indeed very different. The most dramatic change is probably the behavior of the conductivity.

When the perturbations of the metric were overlooked, we, in agreement with some of the studies mentioned in the bibliography, had found that the conductivity in the direction perpendicular to the magnetic field was essentially unperturbed by its presence regardless of its magnitude. Here, as noticed in the previous section, we see that this conductivity drops to zero for any non vanishing background magnetic field. Since the conductivity accounts for how large is the current in a medium induced by an external electric field $E$, we believe this behavior to be more intuitive. We base this believe in the fact that the force acting on the conducting charges due to the background field $B$ for an external electric field $E$ perpendicular to it, will point in the orthogonal direction to $E$, thus reducing the conductivity.

Concerning the conductivity along $B$, we found before that its response was linear in the intensity of $B$. Here, we see that this is still the case for large values of $B$, but there are relevant correction for values of $b/T^2 < 4$. This in particular, read from Fig. \eqref{condZ}, seams to indicate that $\dfrac{d \sigma}{d B}$ vanishes close to $B=0$.

In general the impact of consistently considering the perturbations of the metric is a lot more significant for low frequencies, in particular the conductivity being the zero frequency limit. The reason for these corrections to be of particular importance is that the experimental observations of photon enhancement are relevant for low $P_T$ photons, which is exactly where we are appreciating the impact of the metric perturbations to be stronger.

The statement of the previous paragraph is supported by the results shown in Fig. \eqref{comp}, where we see that for large $\wn$ the spectral densities for both polarizations of the photons propagating in the $x$ direction coincides, and as can be seen in Fig. \eqref{chiyyZ} it does not differ by much from the one for photons propagating in the $z$ direction. Contrarily, for low $\wn$, the effect of the magnetic field is opposite for photons with different polarization, as seen in Fig. \eqref{comp}.

The construction of the family of backgrounds done here, also permits us to see how the discontinuous change in the conductivity arises from a smooth deformation of the spectral density function. In either Fig. \eqref{chiyyZ} or \eqref{chiyyX}, we see how the spectral function is smoothly deformed from its $B=0$ profile as $B$ increases, and how even if the deformation is smooth, as soon as $B$ is different from zero, the perpendicular conductivity vanishes. As $B$ increases, we see how the frequency where the most intense photon production occurs also increases and for photons propagating in the $z$ direction in Fig. \eqref{chiyyZ} the highest value of this peak reduces while for photons propagating in the $x$ direction \eqref{chiyyX} increases.

The shifting of the frequency of highest production and the modification of the maximum density contribute differently to the total production, and that is why the plots of the total photon production are not monotonically ordered as $b/T^2$ goes from 0 to 5. For larger values of $b/T^2$ the enhancement of photon production in the $x$ direction increases with $b/T^2$, while the reverse effect is observed for photons propagating in the $z$ direction.

In a light analysis, we find this general behavior consistent with the physics of the plasma. For photons propagating in the direction of the background magnetic field, the vibrations of the charge bearers have to take place perpendicular to it. To execute low frequencies, larger displacements would have to occur, and these are suppressed by the perpendicular Lorentz force caused by the magnetic field acting on the charge bearers. For larger frequencies this effect would have a smaller impact since the necessary displacements would have to be smaller. For photons propagating in the direction perpendicular to the background field, the same would be true for low frequency photons (with electric field) polarized perpendicular to $B$, but those polarized parallel to $B$ will have no such Lorentz force acting on them. Our results indicate that the enhancement effect should come from a more subtle phenomenon.


\section{Discussion}
\label{sec4}

In this paper we have computed the photon production rate and the electric conductivity of a ${\cal N}=4$ plasma in the presence of an arbitrarily intense magnetic field modelled holographically using a family of solutions that we constructed following \cite{D'Hoker:2009mm}. The plasma is infinitely extended, in thermal equilibrium, strongly coupled and with a large number of colors, though the inclusion of a constant background magnetic field breaks the spatial rotational invariance from SO(3) to SO(2), making the plasma anisotropic. We choose the magnetic field to point in the $z$-direction. In a heavy-ion collision experiment, the latter would correspond to a direction perpendicular to the beam, dictated by the non centrality of the collision, whilst the $xy$-plane would correspond to a plane containing the beam. To study photon distributions in the directions perpendicular to the beam, we can use the SO(2) symmetry in the $x$-$y$ plane to make this beam coincide with the $y$ axis.

As we have already mentioned, the photon production was studied recently \cite{Patino:2012py} for generic angles in a plasma which was also anisotropic, but with the source of the anisotropy a position-dependent theta-angle in the gauge theory, or, equivalently, a position-dependent axion in the gravity background \cite{Mateos:2011ix,Mateos:2011tv}.

In \cite{Patino:2012py} the question was posed about how universal was the behavior that they reported, i.e. if it depended on the source of anisotropy used. Here we conclude that the source of anisotropy is very relevant as we can see from several differences in the effect that the background magnetic field has on the photon production in comparison to the way in which the anisotropy source of \cite{Mateos:2011ix,Mateos:2011tv} is related to the results in \cite{Patino:2012py}, in particular the following three statements are reversed, or contradict the correspondent statement, for these two sources of anisotropy.

a) Here we report that the sensitivity\footnote{The statement about the sensitivity to the intensity of the background field concerns the observations made above about how the strongest effects are seen for low frequencies, where the spectral functions show significant changes for different intensities of $B$, propagation directions and polarizations. All these differences are smaller for large frequencies.} to the intensity of the source of anisotropy and to the angle between the anisotropic direction and the photon wave vector, decreases as the photon energy gets larger (Fig. \eqref{comp}, \eqref{chiyyZ}), \eqref{chiyyZlargeW} and \eqref{chizzXlargeW})).

For the second statement we remember that both, in \cite{Patino:2012py} and here, we see that in the limit of high photon energy, the spectral function divided by the energy of the photon tends to a constant. 

b) Here we report that this constant is independent of the polarization and mildly sensitive to the direction of propagation of the photon (Fig. \eqref{comp}, \eqref{chiyyZlargeW} and \eqref{chizzXlargeW})).

c) In the presence of the background magnetic field the conductivity in the direction perpendicular to it, is reduced (drops to zero) regardless of the intensity of the source of anisotropy, while the conductivity in the direction of the field increases with the intensity of the source of anisotropy (Fig. \ref{condZ}).

It is important to notice that, as well as in \cite{Patino:2012py}, we observe an enhancement of photon production for high $\mathrm{p}_\mathit{T}$ regardless of the propagation direction. This enhancement is with respect to the case with no background magnetic field (isotropic case), and consists solely in a difference in the constant slope of the spectral function with respect to the energy of the photon at high energies. Nonetheless, the strongest enhancement is reported for photons produced perpendicular to $B$ with energies of the order of the temperature of the plasma. For high intensities of the background magnetic field, this enhancement is large enough to make the total production of photons, {\it i.e} integrated over all directions, larger than in the $B=0$ case.

In particular a) and b) are relevant in the views of the observations reported by ALICE and PHENIX, since a) indicates that both, the enhancement of photon production and the anisotropy of the spectrum will be more significant for low $\mathrm{p}_\mathit{T}$, which is qualitatively consistent with the experiments. Furthermore, b) implies also that the anisotropy in the spectrum should diminish for high $\mathrm{p}_\mathit{T}$, again qualitatively consistent with the experiments.

Given that we worked in a field theory that is not QCD, the comparison with experimental data can only be done to the qualitative level that has been discussed here.

Other than the study of generic directions of propagation, which is already being undertaken, there are, at least, two improvements to our calculations which implementation is understood. One is to consider matter in the fundamental representation, and the other is to make this matter fields to have a non vanishing mass\footnote{In a strongly coupled plasma there are no quasi-particles and the mass mentioned here would in fact simply be a microscopic parameter on which the physics depends, but without the direct interpretation of mass of a quasi-particle.}. Both these changes can be performed by using the ten dimensional lift of the five dimensional theory used here and embedding some flavour D7 probe branes on it. The mere inclusion of the D7 branes would add matter in the fundamental representation, while if their embedding is not equatorial, it would also give non vanishing mass to this fields.

The mass of the matter fields having an effect on the photon production rate has already been reported in previously studied backgrounds \cite{Mateos:2007yp,Wu:2013qja}, so it is clear that such a dependence should also be expected in the presence of a background field. In particular, as the mass becomes large compared to the temperature, it should be possible to identify the appearance of highly localized resonances in the spectral functions, indicating the reconstitution of the mesons that were melted in the plasma \cite{Mateos:2007yp}. Understanding the role played by the background magnetic field in this transition is certainly something worth exploring.  

It is worth commenting that the inclusion of massive quarks would be important to make a quantitative estimation of the light produced in quark matter stars, which is another system where the  photon production enhancement by a strong magnetic field has been suggested \cite{MorenoProc} as the relevant mechanism to make contact with possible observations. Quark matter has high plasma frequencies $\simeq$ 20 MeV, thus a compact star made of such material all the way up to its surface would be unable to radiate its thermal emission. In \cite{MorenoProc}, the MIT-bag-model was used to study the propagation modes of photons in such objects when there exists a strong magnetic field, B$>10^{12}$ G. They showed that for an intense enough background field, small windows open in the frequency spectrum which could make them visible. A key ingredient in \cite{MorenoProc} to predict the possibility of the emission of thermal photons in the visual part of the spectrum by these stellar objects is the decrease of the Fermi energy achieved by the inclusion of strange quarks. To first order  approximation, in the bag model, quarks can be assumed massless, so the way in which strange quarks lower the Fermi energy of the system is by making it necessary to take into consideration their non vanishing mass. The results of the present work, particularly a) and c) above, qualitatively point in favor of the possibility of visible photons being emitted by quark matter stars and some quantitative comparison is being prepared \cite{MorenoPatino}, nonetheless the study of photon production by massive quarks in the presence of a background magnetic field is interesting on its own right and such calculation will be carried elsewhere \cite{MorenoPatino2}.

It is imaginable how to implement another improvement regarding the fact that we see that two different sources for anisotropy have had different qualitative effects on the photon production. This would consist in finding a fully back reacted geometry that is dual to a field theory in the presence of a intense background magnetic field along with another source of anisotropy, potentially the one in \cite{Mateos:2011ix,Mateos:2011tv}.

Such hypothetical background would be dual to a theory where all the spacial SO(3) symmetry has been broken, since, to make contact with collision experiments, the background magnetic field would point in a direction perpendicular to the beam, while the other source of anisotropy should single out the direction of the beam itself, somehow modelling the anisotropy coming from the high momentum density concentrated in this direction. Results have been obtained in this direction \cite{Wu:2013qja} without considering the full back reaction of the ten dimensional geometry to the presence of the field nor the perturbations of the embedding of the D7 due to the perturbations of the electromagnetic field necessary to study photon production.

Even after all this improvements we would still not be working in QCD, but the resulting theory would be closer to model the QGP produced in collision experiments. It might be then sensible to perform a more quantitative analysis, involving the comparison of the exponential slope for low $\mathrm{p}_\mathit{T}$ of our model against the experimental measurements.

To the best of our knowledge, so far the detection of direct photons has been done for an integrated rapidity from zero to .35, which corresponds to adding the measurements for angles between zero and approximately 20 degrees around the transverse direction. ALICE has the capability of measuring photon production separately for different rapidities, and also, further away from the zenith, so this refined measurements are planed to be taken in the future. Given that our results show a sizeable difference of photon enhancement depending on the direction of emission, once the experiments that account for small solid angles of detection are performed, some quantitative comparison can be made that should help us understand the origin of the enhancement in photon production observed in this system.


\section*{Acknowledgements}

We are supported by DGAPA-UNAM Grant No. IN117012 (GA, FN and LP). FN acknowledges support from DGAPA-UNAM (postdoctoral fellowship) and CONACYT Grant No. 166391 and DGAPA-UNAM Grant No. IN106110. We would like to thank Gary Horowitz, for his careful reading of the first version of this paper, and for bringing to our attention the profound relevance that in this case has the back reaction generated by the perturbations of the electromagnetic field. We are thankful to  Enrique Moreno for pointing out the relation of the present work to the Quark Matter stars, which was an initial motivation for this study. LP wants to thank David d'Enterria for a very clear explanation about some relevant practical details of the experimental designs to study heavy ion collisions. PO wants to thank Eduardo Barrios for his help while developing some of the codes needed for the early calculations in this work. PO acknowledges support from CONACYT scholarship.


\section{Appendix}

The equations of motion for the fields $A$ and $h$ decouple in two sets, and after imposing $h_{\mu r} = 0$ these two are further simplified. The first one contains the fields $A_t$, $A_x$ and $A_z$, 

\begin{multline}  \label{eqAt} 
U(r)V^2(r)W(r)\ A''_t(r) + U(r)V(r) \bigg[W(r)V'(r) + \frac{1}{2}\ V(r)W'(r) \bigg]\ A'_t(r)\\
 - V(r) \big[k_z^2\ V(r) + k_x^2\ W(r) \big]\ A_t(r) + k_t\ k_x\ V(r)W(r)\ A_x(r)\\
  + k_t\ k_z\ V^2(r)\ A_z(r) - i\ B\ k_x\ W(r)\ h_{ty}(r) = 0,
\end{multline} 

\begin{multline}    \label{eqAx}
U^2(r)V^2(r)W(r)\ A''_x(r) + U(r)V(r) \big[W(r)U'(r) + \frac{1}{2}\ U(r)W'(r) \big]\ A'_x(r) \\
 - k_t\ k_x\ V(r)W(r)\ A_t(r)  + V(r) \big[k_t^2\ W(r) - k_z^2\ U(r) \big]\ A_x(r) \\
  + k_x\ k_z\ U(r)V(r)\ A_z(r)  + i\ B\  \big[ k_z\ U(r)\ h_{yz}(r) - k_t\ W(r)\ h_{ty}(r) \big]  = 0,
\end{multline}

\begin{multline}   \label{eqAz}
U^2(r)V^2(r)W(r)\ A''_z(r) + U(r)V(r) \bigg[ W(r) \Big(U(r)V(r)\Big)' - \frac{1}{2}\ U(r)V(r)W'(r) \bigg]\ A'_z(r) \\
- k_t\ k_z\ V^2(r)W(r)\ A_t(r) + k_x\ k_z\ U(r)V(r)W(r)\ A_x(r) \\ 
+ V(r)W(r) \big[k_t^2\ V(r) - k_x^2\ U(r) \big]\ A_z(r) - i\ B\ k_x\ U(r)W(r)\ h_{yz}(r) = 0,
\end{multline}

\begin{equation}   \label{eqAr}
 k_t\, V(r)W(r)\, A'_t(r) - k_x\, U(r)W(r)\, A'_x(r) - k_z\, U(r)V(r)\, A'_z(r)   = 0,
\end{equation}

\begin{multline}   \label{eqty}
U(r)V^2(r)W(r)\ h''_{ty}(r) + \frac{1}{2}\ U(r)V^2(r)W'(r)\ h'_{ty}(r) \\
- \bigg[ k_z^2\ V^2(r) + \frac{4}{3}\ B^2\ W(r) + k_x^2\ V(r)W(r) + 8\ V^2(r)W(r) - V(r)W(r)U'(r)V'(r) \bigg]\ h_{ty}(r) \\
+ k_t\ V(r) \Big[ k_x\ W(r)\ h_{xy}(r) +  k_z\ V(r)\ h_{yz}(r) \Big] + 4\ i\ B\ V(r)W(r) \Big[ k_x\ A_t(r) - k_t\ A_x(r) \Big] = 0,
\end{multline}

\begin{multline}   \label{eqxy}
U^2(r)V^2(r)W(r)\ h''_{xy}(r) + U(r)V(r) \bigg[ V(r)W(r)U'(r) - U(r)W(r)V'(r) \\ 
+ \frac{1}{2}\ U(r)V(r)W'(r) \bigg]\ h'_{xy}(r) -  k_x\ V^2(r) \Big[ k_t\ W(r)\ h_{ty}(r) - k_z\ U(r)\ h_{yz}(r) \Big] \\
- \bigg[ V^2(r) \Big( k_z^2\ U(r) - k_t^2\ W(r) \Big) - \frac{8}{3}\ B^2\ U(r)W(r) \\
+ 8\ U(r)V^2(r)W(r) - U^2(r)W(r)V'^2(r) \bigg]\ h_{xy}(r) = 0,
\end{multline}

\begin{multline}  \label{eqyz}
U^2(r)V^2(r)W(r)\ h''_{yz}(r) + U(r)V^2(r) \bigg[ W(r)U'(r) - \frac{1}{2}\ U(r)W'(r) \bigg]\ h'_{yz}(r) \\
- k_z\ V(r)W(r) \Big[ k_t\ V(r)\ h_{ty}(r) - k_x\ U(r)\ h_{xy}(r) \Big] + \bigg[ V(r)W(r) \Big(k_t^2\ V(r) - k_x^2\ U(r) \Big) \\
- \frac{4}{3}\ B^2\ U(r)W(r) - 8\ U(r)V^2(r)W(r) + U^2(r)V(r)V'(r)W'(r) \bigg]\ h_{yz}(r) \\
+ 4\ i\ B\ U(r)V(r)W(r) \Big[k_x\ A_z(r) - k_z\ A_x(r) \Big] = 0,
\end{multline}

\begin{multline}   \label{eqyr}
V(r) \Big[ k_t\ V(r)W(r)\ h'_{ty}(r) - k_x\ U(r)W(r)\ h'_{xy}(r) \\ 
- k_z\ U(r)V(r)\ h'_{yz}(r) \Big] + 4\ i\ B\ U(r)V(r)W(r)\ A'_x(r) \\ 
- V'(r) \Big[ k_t\ V(r)W(r)\ h_{ty}(r) - k_x\ U(r)W(r)\ h_{xy}(r) - k_z\ U(r)V(r)\ h_{yz}(r) \Big] = 0,
\end{multline}

The second set contains the field $A_y$, 

\begin{multline}  \label{eqAy}
U^2(r)V^2(r)W(r)\ A''_y(r) + U(r)V^2(r) \bigg[ W(r)U'(r) + \frac{1}{2}\ U(r)W'(r) \bigg]\ A'_y(r) \\
+ V(r) \Big[ k_t^2\ V(r)W(r) - k_x^2\ U(r)W(r) - k_z^2\ U(r)V(r) \Big]\ A_y(r) \\
+ \frac{1}{2}\ i\ B\ k_x \bigg[ V(r)W(r)\ h_{tt}(r) - U(r)W(r)\ h_{xx}(r) - U(r)W(r)\ h_{yy}(r) + U(r)V(r)\ h_{zz}(r) \bigg] \\
+ i\ B\ V(r) \Big[ k_t\ W(r)\ h_{tx}(r) - k_z\ U(r)\ h_{xz}(r) \Big] = 0,
\end{multline}

\begin{multline}   \label{eqtt}
U^2(r)V^3(r)W^2(r)\ h''_{tt}(r) + U(r)V^2(r)W(r) \bigg[ U(r)W(r)V'(r) + \frac{1}{2}\ U(r)V(r)W'(r) \\
- \frac{1}{2}\ V(r)W(r)U'(r) \bigg]\ h'_{tt}(r) + \frac{1}{2}\ U^2(r)V^2(r)W(r)U'(r) \Big[ W(r) \Big( h'_{xx}(r) + h'_{yy}(r) \Big) + V(r)\ h'_{zz}(r) \Big] \\
- V(r)W(r) \bigg[ U(r)V(r) \Big( k_z^2\ V(r) + k_x^2\ W(r) \Big) + \frac{4}{3}\ B^2\ U(r)W(r) + 8\ U(r)V^2(r)W(r) \\ 
- 3\ V^2(r)W(r)U'^2(r) \bigg]\ h_{tt}(r) + U(r)W^2(r) \bigg[\frac{4}{3}\ B^2\ U(r) + k_t^2\ V^2(r)\\ 
- \frac{1}{2}\ U(r)V(r)U'(r)V'(r) \bigg] \Big( h_{xx}(r) + h_{yy}(r) \Big) + U(r)V^3(r) \bigg[ k_t^2\ W(r) - \frac{1}{2}\ U(r)U'(r)W'(r) \bigg]\ h_{zz}(r) \\
 - 2\ k_t\ U(r)V^2(r)W(r) \Big[ k_z\ V(r)\ h_{tz}(r) + k_x\ W(r)\ h_{tx}(r) \Big] \\ 
 - \frac{8}{3}\ i\ B\ k_x\ U^2(r)V(r)W^2(r)\ A_y(r) = 0, 
\end{multline}

\begin{multline} \label{eqtx}
U(r)V^2(r)W(r)\ h''_{tx}(r) + \frac{1}{2}\ U(r)V^2(r)W'(r)\ h'_{tx}(r) - k_t\ k_x\ V(r) \Big[ W(r)\ h_{yy}(r) + V(r)\ h_{zz}(r) \Big] \\
- \bigg[ k_z^2\ V^2(r) + \frac{4}{3}\ B^2\ W(r) + 8\ V^2(r)W(r) - V(r)W(r)U'(r)V'(r) \bigg]\ h_{tx}(r) \\
+ k_z\ V^2(r) \Big[ k_x\ h_{tz}(r) + k_t\ h_{xz}(r) \Big] + 4\ i\ B\ k_t\ V(r)W(r)\ A_y(r) = 0,
\end{multline}

\begin{multline}  \label{eqtz}
U(r)V^2(r)W(r)\ h''_{tz}(r) - U(r)V(r) \bigg[W(r)V'(r) - \frac{1}{2}\ V(r)W'(r) \bigg]\ h'_{tz}(r) \\
- k_t\ k_z\ V(r)W(r) \Big[ h_{xx}(r) + h_{yy}(r) \Big] + k_x\ V(r)W(r) \Big[ k_z\ h_{tx}(r) + k_t\ h_{xz}(r) \Big] \\
- \bigg[ \frac{4}{3}\ B^2\ W(r) + k_x^2\ V(r)W(r) + 8\ V^2(r)W(r) - V^2(r)U'(r)W'(r) \bigg] = 0,
\end{multline}

\begin{multline}  \label{eqtr}
2\ k_t\ U(r)V(r)W(r) \Big[ W(r)\ h'_{xx}(r) + W(r)\ h'_{yy}(r) + V(r)\ h'_{zz}(r) \Big] \\
-2\ U(r)V(r)W(r) \Big[ k_x\ W(r)\ h'_{tx}(r) + k_z\ V(r)\ h'_{tz}(r) \Big] \\
 +  k_t\ W^2(r) \Big[ V(r)U'(r) - U(r)V'(r) \Big] \Big( h_{xx}(r) - h_{yy}(r) \Big) - k_t\ V^2(r) \Big[ W(r)U'(r) + U(r)W'(r)\Big]\ h_{zz}(r) \\ 
 + 2\ V(r)W(r)U'(r) \Big[ k_z\ V(r)\ h_{tz}(r) + k_x\ W(r)\ h_{tx}(r) \Big] = 0,
\end{multline}

\begin{multline}   \label{eqxx}
U^2(r)V^2(r)W^2(r)\ h''_{xx}(r) \\ 
+ \frac{1}{2}\ U(r)V(r)W(r) \bigg[ 2\ V(r)W(r)U'(r) - U(r)W(r)V'(r) + U(r)V(r)W'(r) \bigg]\ h'_{xx}(r) \\
 + \frac{1}{2}\ U(r)V(r)W(r) \bigg[ V(r)W(r)\ h'_{tt}(r) + U(r)W(r)\ h'_{yy}(r) + U(r)V(r)V'(r)\ h'_{zz}(r) \bigg] \\
- V^2(r)W^2(r) \bigg[ k_x^2\ + \frac{1}{2}\ U'(r)V'(r) \bigg]\ h_{tt}(r) - U(r)V^2(r) \bigg[ k_x^2\ W(r) + \frac{1}{2}\ U(r)V'(r)W'(r) \bigg]\ h_{zz}(r) \\
- W(r) \bigg[ k_z^2\ U(r)V^2(r) - k_t^2\ V^2(r)W(r) + 8\ U(r)V^2(r)W(r) - \frac{1}{2}\  U^2(r)W(r)V'^2(r) \bigg]\ h_{xx}(r) \\
- U(r)W^2(r) \bigg[ \frac{8}{3}\ B^2 + k_x^2\ V(r) + \frac{1}{2}\ U(r)V'^2(r) \bigg]\ h_{yy}(r) \\ 
+ 2\ k_x\ V^2(r)W(r) \Big[ k_z\ U(r)\ h_{xz}(r) U(r) - k_t\ W(r)\ h_{tx}(r) \Big] \\
+ \frac{16}{3}\ i\ B\ k_x\ U(r)V(r)W^2(r)\ A_y(r) = 0,
\end{multline}

\begin{multline}  \label{eqxz}
U^2(r)V^2(r)W(r)\ h''_{xz}(r) + U(r)V^2(r) \bigg[ W(r)U'(r) - \frac{1}{2}\ U(r)W'(r) \bigg]\ h'_{xz}(r) \\
- k_x\ k_z\ V(r)W(r) \Big[ V(r)\ h_{tt}(r) + U(r)\ h_{yy}(r) \Big] - k_t\ V^2(r)W(r) \Big[ k_z\ h_{tx}(r) + k_x\ h_{tz}(r) \Big] \\
- \bigg[ \frac{4}{3}\ B^2\ U(r)W(r) - k_t^2\ V^2(r)W(r) + 8\ U(r)V^2(r)W(r) - U^2(r)V(r)V'(r)W'(r) \bigg]\ h_{xz}(r) \\
+ 4\ i\ B\ k_z\ U(r)V(r)W(r)\ A_y(r) = 0,
\end{multline}

\begin{multline}  \label{eqxr}
k_x\ U(r)V(r)W(r) \Big[ V(r)W(r)\ h'_{tt}(r) + U(r)W(r)\ h'_{yy}(r) + U(r)V(r)\ h'_{zz}(r) \Big] \\
+ U(r)V^2(r)W(r) \Big[ k_t\ W(r)\ h'_{tx}(r) - k_z\ U(r)\ h'_{xz}(r) \Big] - 4\ i\ B\ U^2(r)V(r)W^2(r)\ A'_y(r) \\
- \frac{1}{2}\ k_x\ V(r)W^2(r) \Big[U(r)V(r)\Big]'\ h_{tt}(r) - \frac{1}{2}\ k_x\ U^2(r)V(r) \Big[V(r)W(r)\Big]'\ h_{zz}(r) \\
- k_x\ U^2(r)V^2(r)W'(r) + U(r)V(r)W(r)V'(r) \Big[ k_z\ U(r)\ h_{xz}(r) - k_t\ W(r)\ h_{tx}(r) \Big] = 0,
\end{multline}

\begin{multline}  \label{eqyy}
U^2(r)V^2(r)W^2(r)\ h''_{yy}(r) \\
 + \frac{1}{2}\ U(r)V(r)W(r) \bigg[ 2\ V(r)W(r)U'(r) - U(r)W(r)V'(r) + U(r)V(r)W'(r) \bigg]\ h'_{yy}(r) \\
\frac{1}{2}\ U(r)V(r)W(r)V'(r) \bigg[V(r)W(r)\ h'_{tt}(r) + U(r)W(r)\ h'_{xx}(r) + U(r)V(r)\ h'_{zz}(r) \bigg] \\
-\frac{1}{2}\ V^2(r)W^2(r)U'(r)V'(r)\ h_{tt}(r) - U(r)W^2(r) \bigg[ \frac{8}{3}\ B^2 + \frac{1}{2}\ U(r)V'^2(r) \bigg]\ h_{xx}(r) \\
-  W(r) \bigg[ V(r) \Big( k_z^2\ U(r)V(r) + k_x^2\ U(r)W(r) - k_t^2\ V(r)W(r) \Big) \\ 
+ 8\ U(r)V^2(r)W(r) - U^2(r)W(r)V'^2(r) \bigg]\ h_{yy}(r) \\
- \frac{1}{2}\ U^2(r)V^2(r)V'(r)W'(r)\ h_{zz}(r) + \frac{16}{3}\ i\ B\ k_x\ U(r)V(r)W^2(r)\ A_y(r) = 0,
\end{multline}

\begin{multline}  \label{eqzz}
U^2(r)V^3(r)W^2(r) \ h''_{zz}(r) + \frac{1}{2}\ U(r)V^2(r)W^2(r)W'(r) \bigg[ V(r)\ h'_{tt}(r) + U(r)\ h'_{xx}(r) + U(r)\ h'_{yy}(r) \bigg] \\
+ U(r)V^2(r)W(r) \Big[ V(r)W(r)U'(r) + U(r)W(r)V'(r) + U(r)V(r)W'(r) \Big]\ h'_{zz}(r) \\ 
- \frac{1}{2}\ V^3(r)W^2(r) \bigg[ 2\ k_z^2 + U'(r)W'(r) \bigg]\ h_{tt}(r) \\
- U(r)W^2(r) \bigg[ k_z^2\ V(r) - \frac{4}{3}\ B^2\ W(r) + \frac{1}{2}\ U(r)V(r)V'(r)W'(r) \bigg] \Big( h_{xx} + h_{yy} \Big) \\
- V(r) \bigg[ \frac{4}{3}\ B^2\ U(r)W^2(r) + 6\ V(r)W^2(r) \Big( k_x^2\ U(r)  - k_t^2\ V(r) \Big) + 8\ U(r)V^2(r)W^2(r) \\ 
- \frac{1}{2}\ U^2(r)V^2(r)W'(r) \bigg]\ h_{zz}(r)  + 2\ k_z\ V^2(r)W^2(r) \Big[ k_x\ U(r)\ h_{xz}(r) - k_t\ V(r) h_{tz}(r) \Big]  \\
- \frac{8}{3}\ i\ B\ k_x\ U(r)V(r)W^3(r)\ A_y(r) = 0,
\end{multline}

\begin{multline}  \label{eqzr}
k_z\ U(r)V(r)W(r) \Big[ V(r)\ h'_{tt}(r) + U(r)\ h'_{xx}(r) + U(r)\ h'_{yy}(r) \Big] \\
 + U(r)V(r)W(r) \Big[ k_t\ V(r)\ h'_{tz}(r) - k_x\ U(r)\ h'_{xz}(r) \Big] \\
- \frac{1}{2}\ k_z\ V^2(r) \Big[U(r)W(r)\Big]'\ h_{tt}(r) - \frac{1}{2}\ k_z\ U^2(r) \Big[V(r)W(r)\Big]' \Big(h_{xx}(r) + h_{yy}(r) \Big) \\
+ U(r)V(r)W'(r) \Big[ k_x\ U(r)\ h_{xz}(r) - k_t\ V(r)\ h_{tz}(r) \Big] = 0,
\end{multline} 

\begin{multline}  \label{eqrr}
U^2(r)V^2(r)W^2(r) \Big[ V(r)W(r)\ h''_{tt}(r) + U(r)W(r)\ \Big( h''_{xx}(r) + h''_{yy}(r) \Big) + U(r)V(r)\ h''_{zz}(r) \Big] \\
- \frac{1}{2}\ U(r)V^3(r)W^3(r)U'(r)\ h'_{tt}(r) + \frac{1}{2}\ U^2(r)V(r)W^3(r) \bigg[ V(r)U'(r) - 2\ U(r)V'(r) \bigg] \Big( h'_{xx}(r) + h'_{yy}(r) \Big) \\
+ \frac{1}{2}\ U^2(r)V^3(r)W(r) \Big(W(r)U'(r) - 2\ U(r)W'(r) \Big)\ h'_{zz}(r) + \frac{1}{2}\ V^3(r)W^3(r) \bigg[ U'^2(r) - 2\ U(r)U''(r) \bigg] \\
+ U^2(r)W^3(r) \bigg[ \frac{4}{3}\ B^2 - \frac{1}{2}\ V(r)U'(r)V'(r) + U(r)V'^2(r) - U(r)V(r)V''(r) \bigg] \Big( h_{xx}(r) + h_{yy}(r) \Big) \\
- U^2(r)V^3(r) \bigg[ U(r)W'^2(r) - U(r)W(r)W''(r) - \frac{1}{2}\ W(r)U'(r)W'(r) \bigg]\ h_{zz}(r)\\
- \frac{8}{3}\ i\ B\ k_z\ U^2(r)V(r)W^3(r) = 0,
\end{multline}


\end{document}